\documentclass[sn-mathphys,Numbered]{sn-jnl}

\usepackage{graphicx}%
\usepackage{multirow}%
\usepackage{amsmath,amssymb,amsfonts}%
\usepackage{amsthm}%
\usepackage{mathrsfs}%
\usepackage[title]{appendix}%
\usepackage{xcolor}%
\usepackage{textcomp}%
\usepackage{manyfoot}%
\usepackage{booktabs}%
\usepackage{algorithm}%
\usepackage{algorithmicx}%
\usepackage{algpseudocode}%
\usepackage{listings}%

\usepackage{subcaption}
\usepackage[euler]{textgreek}
\usepackage[nice]{nicefrac}

\theoremstyle{thmstyleone}%
\theoremstyle{thmstyletwo}%

\theoremstyle{thmstylethree}%

\begin{document}

\title[Article Title]{Article Title}


\title[Time-Reflection of Microwaves by a Fast Optically-Controlled Time-Boundary]{Time-Reflection of Microwaves by a Fast Optically-Controlled Time-Boundary}




\author*[1]{\fnm{Thomas R.} \sur{Jones}}\email{jone2006@purdue.edu}

\author[1]{\fnm{Alexander V.} \sur{Kildishev}}\email{kildishev@purdue.edu}

\author[2]{\fnm{Mordechai} \sur{Segev}}\email{msegev@technion.ac.il}

\author[1]{\fnm{Dimitrios} \sur{Peroulis}}\email{dperouli@purdue.edu}

\affil[1]{\orgdiv{Elmore Family School of Electrical and Computer Engineering}, \orgname{\\* Purdue University}, \city{West Lafayette}, \state{IN}, \country{USA}}

\affil[2]{\orgdiv{Department of Physics}, \orgname{Technion-Israel Institute of Technology},\city{\\* Haifa}, \country{Israel}}


\abstract{When an electromagnetic (EM) wave is propagating in a medium whose properties are varied abruptly in time, the wave experiences refractions and reflections known as ``time-refractions'' and ``time-reflections'', both manifesting spectral translation as a consequence of the abrupt change of the medium and the conservation of momentum. However, while the time-refracted wave continues to propagate with the same wave-vector, the time-reflected wave is propagating backward with a conjugate phase, despite the lack of any spatial interface. Importantly, while time-refraction is always significant, observing time-reflection poses a major challenge -- because it requires a large change in the medium occurring within a single cycle. For that reason, time-reflection of EM waves was observed only recently. Here, we present the observation of microwave pulses at the highest frequency ever observed (0.59~GHz), and the experimental evidence of the phase-conjugation nature of time-reflected waves. Our experiments are carried out in a periodically-loaded microstrip line with optically-controlled picosecond-switchable photodiodes. Our system paves the way to the experimental realization of Photonic Time-Crystals at GHz frequencies.}






\maketitle

\section{Introduction}\label{sec1}

When the refractive index is changed abruptly in time, a wave propagating in the medium experiences time-refraction and time-reflection, similar to the refraction and reflection at dielectric interfaces \cite{Mendonca_2002,Zhou_2020,Galiffi_2022}. However, despite the similarity, time-reflection and refraction are fundamentally different than their spatial counterparts. Namely, whereas energy (frequency) is conserved at spatial interfaces between dielectric media, a time-interface necessarily changes the frequency, because time-translation symmetry is broken by the temporal modulation. On the other hand, when the abrupt temporal change occurs in a homogeneous medium - momentum (wave-vector $k$) is conserved, which implies that both the time-refraction and the time-reflection experience translation of their spectrum \cite{Morgenthaler_1958,Shaltout_2015}. Also, causality implies that the time-reflection cannot go back in time (unfortunately) but instead is back-reflected in space. That is, while the time-refracted wave continues to propagate with the same wave-vector, the time-reflected wave is propagating backwards with a conjugate phase \cite{Lerosey_2004, Biancalana_2007}. In this way, modulating the refractive index strongly and periodically in time gives rise to multiple time-reflections and time-refractions, which interfere and yield dispersion bands separated by band-gaps in the momentum $k$ \cite{Zurita_2012,Shaltout2_2016,Lustig_2018}. This temporal structure is known as Photonic Time-Crystals (PTCs), and their most intriguing property is that states residing in the momentum gap can have exponentially increasing amplitudes, which draw energy from the modulation \cite{Holberg_1966,Zurita_2009,Lyubarov_2022}. Importantly, PTCs offer a plethora of new possibilities of new physics and applications, ranging from the generation of pairs of entangled photons and below-threshold Cherenkov radiation to new widely-tunable laser sources at THz frequencies \cite{Shaltout_2016,Shlivinski_2018,Gaxiola_2021,Sharabi_2022,Apffel_2022,Jaffray_2022,Lustig_2023,Liu_2023}. However, all of these rely on time-reflections: without significant time-reflections there would be no PTCs. This understanding poses a major challenge: for the time-reflection to be significant, the refractive index change has to be large (order of unity) and abrupt (occurring within 1-2 cycles of the time-reflected waves), otherwise the time-reflection is extremely weak, and PTCs become impossible \cite{Saha_2023,Lustig2_2023}. This tough requirement to have a strong and abrupt change in the refractive index is the reason why the universal phenomenon of time-reflected waves has been observed only recently. Its properties were thus far studied with water waves \cite{Bacot_2016} and with EM waves at radio frequencies \cite{Reyes_2015}, and only very recently with microwaves \cite{Moussa_2023}. 

Here, we study the propagation of microwaves in a periodically-loaded microstrip line with optically-controlled picosecond-switchable photodiodes, and present time-reflection at the highest frequency ever observed (0.59~GHz) along with direct experimental evidence for its phase-conjugate nature. Our experiments demonstrate the time-resolved reflection of an ultra-high frequency pulse propagating along a periodically-loaded microstrip transmission line, occurring due to a time-modulated distributed impedance.

For a strong time-reflection at microwave frequencies, the change in the characteristic impedance due to the temporal event must be fast – within 1-2 cycles of the time-reflected waves. Consequently, we use high-speed photodiodes with switching times in the picosecond regime, to control the capacitive loading and achieve the desired bandwidth. Moreover, the entire experimental setup is judiciously designed to maximize the available bandwidth in both states, such that we can study the propagation of very short pulses in the microstrip line. Also, the electric length of the microstrip line is optimized to be as short as possible, so that the pulse always remains completely within the line during the temporal event. Such a unique experimental system allows us to explore the evolution of ultra high frequency signals, and observe time-reflections at 0.59~GHz occurring due to an almost-instantaneous time-interface. Subsequently, we study the propagation of a hybrid two-peak asymmetric pulse, and demonstrate the time-reversal nature of the time-reflected signal, along with its frequency translation arising from conservation of momentum. We also study the time reflection of a chirped pulse with phase variation, providing experimental evidence of the phase-conjugation nature of time-reflected waves.

Apart from the exciting fundamental aspects of this study, our prototype system has important practical value. It employs realistic microwave circuit designs with surface-mount high-speed photodiodes, and demonstrates the viable potential of ultra-fast time-modulated microstrip circuits for nascent real-world applications in novel microwave pre- and post-processing systems, AI-driven navigation, and satellite communication platforms.

The paper is organized as follows. First, Section~2 describes the general concept of our time-boundary experiment and provides the theoretical background. Section~3 outlines the design of the time-reconfigurable microstrip-line and switchable capacitive loading using high-speed photodiodes. The experimental setup and results are presented in Section~4. Here the frequency response of the time-reconfigurable microstrip is reported, along with the time-boundary experiments using single, two-peak asymmetric, and chirped pulses demonstrating time-reflection and its time-reversal and phase-conjugate nature, along with spectral translation due to conservation of momentum. The conclusion provides a short summary and discussion on future ideas.

\section{Principle of Operation}\label{sec2}

\begin{figure}[h]
\centering
\includegraphics[width=4.5in]{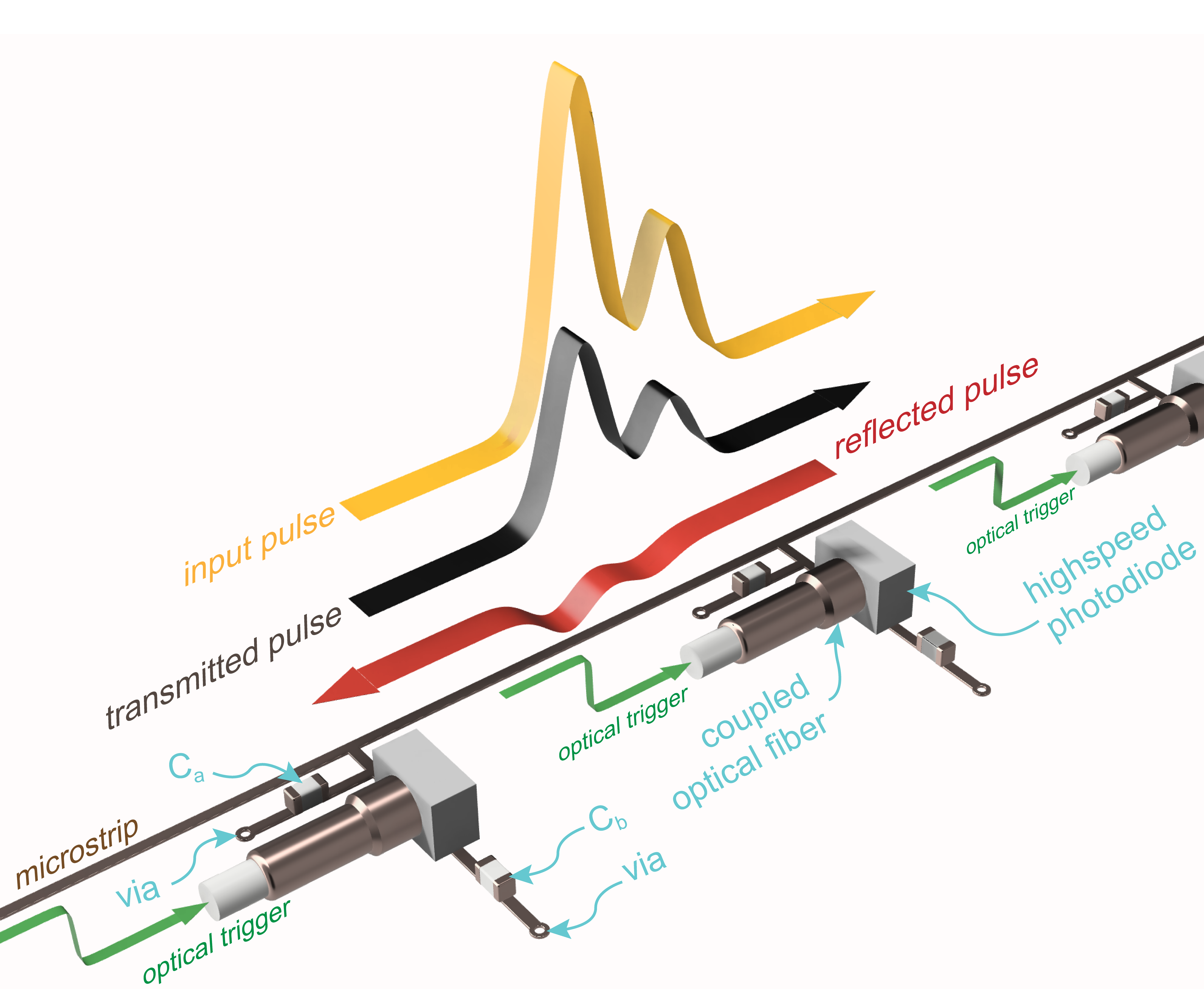}
\caption{\label{fig:3d_concept} \textbf{A conceptual 3D illustration of a time boundary's impact on a two-peak asymmetric pulse.} An input two-peak assymetric pulse (yellow "input pulse") with the smaller amplitude peak leading the larger, propagates along a periodically-loaded microstrip transmission line. The effective impedance of the microstrip is modulated in time by high-speed photodiodes. The diodes are synchronized by simultaneous optical triggers (green "optical trigger") delivered through the optical fibers from a single laser diode (not shown). Upon activation of the time interface, the input pulse is scattered into a transmitted pulse (black "transmitted pulse") and reflected pulse (red "reflected pulse"). The transmitted pulse continues to travel in the same direction as the input with the smaller amplitude peak still leading the larger, with reduced magnitude. The reflected pulse travels in the opposite direction as the input -- the larger amplitude peak now leads the lower one, demonstrating the time-reversal property of PTCs. The pulses are shown in the spatial domain propagating along the x-axis.}
\end{figure}

In this work, a reconfigurable periodically-loaded microstrip with rapidly switchable effective characteristic impedance models an abrupt change in the effective permittivity of the medium. A time interface is generated while a microwave pulse propagates within the microstrip, and the corresponding reflected and transmitted signals are measured at the input and output, respectively. 

Fig.~\ref{fig:3d_concept} illustrates a 3D concept of our reconfigurable distributed microwave transmission line (DMTL) used to experimentally demonstrate time-boundary reflection and transmission of a microwave pulse. The DMTL is designed to be periodically loaded with two parallel shunt capacitors, $C_\mathrm{a}$ and $C_\mathrm{b}$. The second capacitor $\left( C_\mathrm{b}\right)$ is connected in series to a switch controlled by a high-speed photodiode. Both capacitors are terminated with conductive vias to ground. Upon excitation by light (green "optical trigger" pulse), the high-speed photodiodes switch from an isolating OFF-state to a conductive ON-state, connecting the additional shunt capacitor $C_\mathrm{b}$ to the circuit. This reconfiguration changes the effective impedance of the microstrip from State~1 to a lower impedance value in State~2, generating a time-boundary reflection (red "reflected pulse") and transmission (black "transmitted pulse") of the input (yellow "input pulse") wave travelling within the medium. 

With picosecond rise times of high-speed photodiodes, the time interface is much faster than the operating bandwidth of the microwave pulse, enabling a near-instantaneous time-boundary akin to spatial reflections with an abrupt spatial-boundary and allowing the measurement of an ultra high frequency time-reflected pulse of 0.59 GHz. In addition to fast rise times, the use of photodiodes enables almost perfect synchronization of each switch periodically loading the microstrip, in which the optical trigger is produced from a single laser diode source, generating a truly homogeneous time-switched medium.

Two unique properties distinguish time-reflected waves from spatially-reflected waves \cite{Mendonca_2002}. The first is time reversal, due to the homogeneous change in space of the properties of the medium. This instant change leads to a broadband time reversal of the reflected wave. Fig.~\ref{fig:3d_concept} illustrates this effect in the spatial domain with a two-peak asymmetric pulse scattered at a time interface. Observing the input pulse (yellow trace), we see that the higher amplitude peak lags the lower one as the pulse travels towards the top right along the x-axis in the positive x-direction. Upon activation of the time boundary, the order of the peaks is now reversed -- the low-amplitude pulse now lags behind the higher one, as the direction of travel changes towards the bottom left or negative x-direction. This effect is completely different with a spatial boundary -- for a wave bouncing off a spatial discontinuity (spatial wall) the leading edge reaching the wall is the first one to get reflected and received back at the source. With a time boundary, the entire pulse inside the "slab" gets reflected at the same time. This distributed reflection effectively reverses the pulse in time, its trailing edge becomes the leading edge, and the leading edge becomes the trailing one.

The second unique property of time-reflected waves is their broadband frequency translation due to spatial translational symmetry, since the wave's momentum is conserved across the time interface. Following \cite{Moussa_2023}, we derive the frequency translation due to a scattering event at a time interface based on a distributed circuit model of a microwave transmission line. For details of this derivation, see Supplementary~Section~1. The frequency of the input wave $(\omega_1)$ is transformed to that of the reflected and transmitted waves $(\omega_2)$ as,

\begin{equation}
\omega_{2} = \sqrt{\frac{C_{1}}{C_{2}}}\omega_{1} = \frac{Z_{2}}{Z_{1}}\omega_{1} \label{eq_2_1}
\end{equation}

\hfill

\noindent where $C_1$ and $Z_1$ are the loading capacitance and effective characteristic impedance of the DMTL in State~1, respectively, while $C_2$ and $Z_2$ are the loading capacitance and effective characteristic impedance of the DMTL in State~2, respectively. This straightforward model immediately leads to the time-boundary reflection and transmission coefficients: 

\begin{equation}
R = \frac{1}{2}\left[\frac{C_{1}}{C_{2}} - \sqrt{\frac{C_{1}}{C_{2}}}\right] = \frac{Z_{2}(Z_{2} - Z_{1})}{2Z_{1}^2} \label{eq_2_2}
\end{equation}

\begin{equation}
T = \frac{1}{2}\left[\frac{C_{1}}{C_{2}} + \sqrt{\frac{C_{1}}{C_{2}}}\right] = \frac{Z_{2}(Z_{2} + Z_{1})}{2Z_{1}^2} \label{eq_2_3}
\end{equation}

\hfill

The above properties and theoretical derivations of time-boundaries will now be validated through experiments in the following sections.


\section{Design}\label{sec3}

This section now outlines the design of a time-reconfigurable periodically-loaded microstrip line. First, the detailed design steps of a reconfigurable periodically-loaded microstrip line are presented, leading to a realizable prototype suitable for our time-boundary experiment. The details of the time-switched capacitor loading element using high-speed photodiodes are then outlined, with considerations towards the photodiodes measured dc characteristics, synchronization, and switching speed.

\subsection{Time-Reconfigurable Periodically-Loaded Microstrip Line}\label{subsec1}

\begin{figure}[h]
\centering
\includegraphics[width=4.5in]{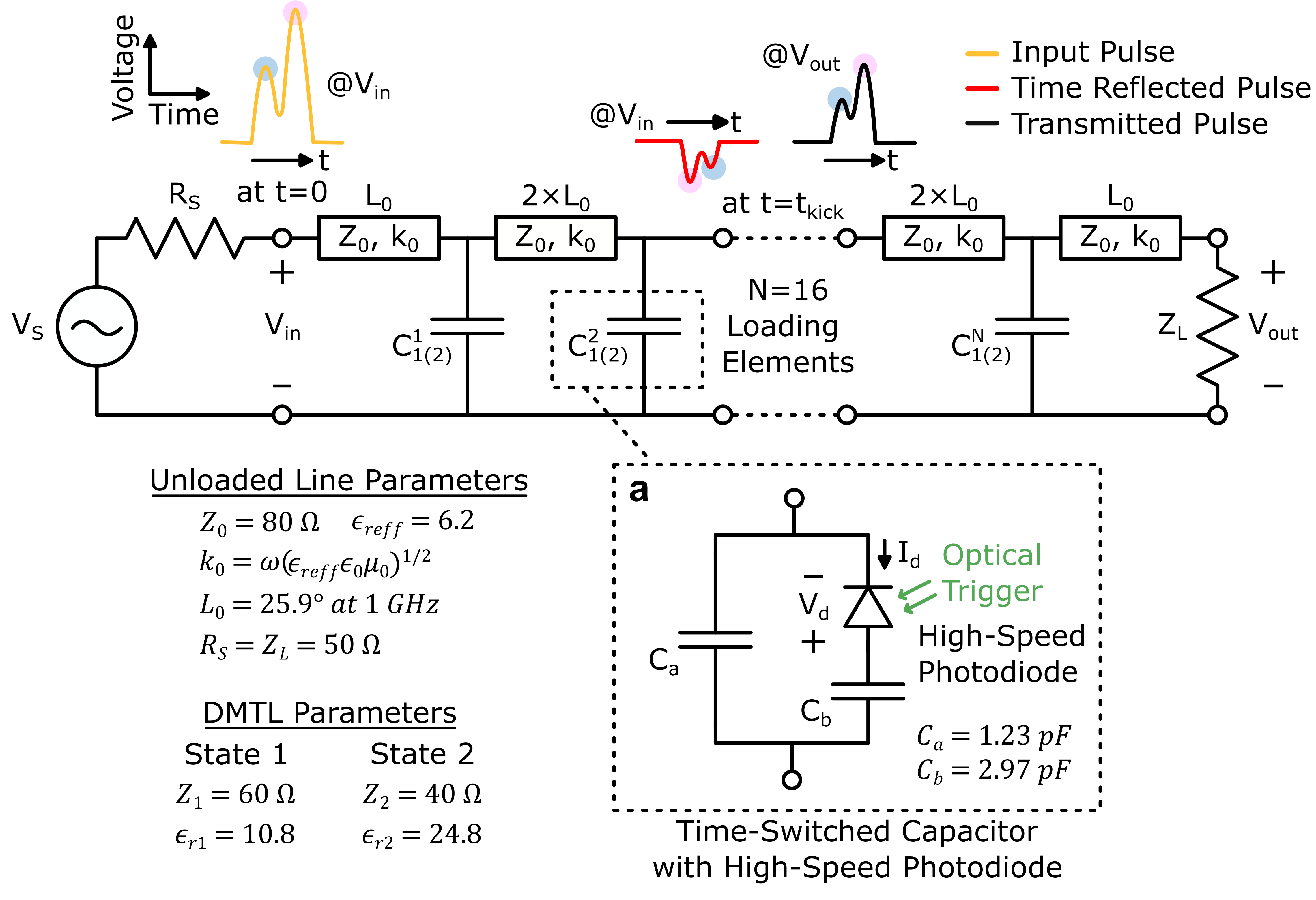}
\caption{\label{fig:equiv_cct} \textbf{The equivalent circuit of a time-reconfigurable periodically-loaded microstrip line. a}, Circuit diagram of a time-switched capacitor, where a high-speed photodiode triggered by an optical pulse changes the impedance of the load from State~1 with $C_1=C_\mathrm{a}$ to State~2 with $C_2=C_\mathrm{a}+ C_\mathrm{b}$. Above the equivalent circuit diagram, the effect of time scattering on a two-peak asymmetric pulse is shown in the time domain, compared to the spatial domain shown in Fig.~\ref{fig:3d_concept}. Here, the input pulse (yellow trace) is measured at voltage reference $V_\mathrm{in}$ versus time, with the lower amplitude peak measured before the higher peak. Upon activation of the time boundary, the transmitted pulse (black trace) measured at voltage reference $V_\mathrm{out}$ maintains the same pulse profile as the input. However, the reflected pulse (red trace) measured at $V_\mathrm{in}$ becomes reversed in time compared to the input, with the higher amplitude peak measured before the lower one.}
\end{figure}

The design methodology of the time-reconfigurable periodically-loaded microstrip transmission line follows the procedure for reconfigurable DMTLs in \cite{Vaha_2004,Barker_2000}. Fig.~\ref{fig:equiv_cct} shows the equivalent circuit diagram where each loading element consists of two shunt capacitors $C_\mathrm{a}$ and $C_\mathrm{b}$, along with a high-speed photodiode configured to operate as a switch. The photodiode is connected in series with $C_\mathrm{b}$, and is reversed biased by a dc voltage applied at the input of the transmission line (not shown). With no light incident on the reversed biased photodiode, the switch is in the OFF-state -- the time-switched capacitor is in State~1, with a total loading capacitance of $C_1=C_\mathrm{a}$. When light with optical power $P$ is incident on the photodiode, the switch moves into the ON-state (State 2), and $C_\mathrm{b}$ becomes connected in parallel to $C_\mathrm{a}$. The total loading capacitance in State~2 becomes $C_2=C_\mathrm{a}+C_\mathrm{b}$. Thus, by controlling all photodiodes with synchronized light pulses, the total loading capacitance, or characteristic impedance, of the entire DMTL is reconfigured in time. 

For a DMTL, the characteristic impedances of the loaded line in State~1(2) are calculated as \cite{Vaha_2004,Barker_2000},

\begin{align}
Z_{1(2)} &= \frac{Z_0}{\sqrt{K_{1(2)}}}\\
K_{1(2)} &= 1 + \frac{C_{1(2)}}{d} \frac{c_0 Z_0}{\sqrt{\varepsilon_\mathrm{reff}}}
\end{align}\label{eq_3_1}

\noindent
where $c_0$ is the free-space velocity of light in vacuum, $K_{1(2)}$ is the scaling factor due to the periodic loading in each state, and $C_{1(2)}$ is the capacitance of each loading element, either in State~1 ($C_1 = C_\mathrm{a}$) or State~2 ($C_2 = C_\mathrm{a} + C_\mathrm{b}$), $d$ is the separation distance between loading elements. $Z_0$ and $\varepsilon_\mathrm{reff}$ are the characteristic impedance of the unloaded microstrip and the effective dielectric constant of the unloaded transmission line \cite{Pozar_2012}. The effective dielectric constant of the loaded transmission line in either State~1 or State~2 can then be calculated as,

\begin{equation}
\varepsilon_\mathrm{reff-1(2)}^\mathrm{DMTL} = \varepsilon_\mathrm{reff}K_{1(2)}
\label{eq_3_3}
\end{equation}

\hfill

As the guided wavelength within a periodically-loaded structure approaches the separation distance between loadings, a cutoff frequency called the Bragg frequency $(f_\mathrm{Bragg-1(2)})$ for State 1(2), 

\begin{equation}
f_\mathrm{Bragg-1(2)} = \frac{c_0 Z_{1(2)}}{\pi d Z_0 \sqrt{\varepsilon_\mathrm{reff}}}
\label{eq_3_4}
\end{equation}

\hfill

\noindent Here, due to the smaller impedance of $Z_2$ compared to $Z_1$, the lower Bragg frequency in State~2 limits the overall bandwidth of the DMTL. 

These constraints must be carefully accounted for to maximize the available bandwidth of the line. Further, the loading elements have self-resonance frequencies which can contribute to operating bandwidth limitations; however, these can be also designed to resonate at frequencies above the Bragg frequency, and are therefore less significant.

Within the linear operating region of the microstrip, the phase velocity ${v_\mathrm{p-1(2)} = c_0 (\varepsilon_\mathrm{reff-1(2)}^\mathrm{DMTL})^{-1/2}}$ is equivalent to the group velocity, and is used to determine the time-frame when the pulse is within the transmission line.



In our study, the effective impedance of the microstrip is designed to change from 60~$\Omega$ to 40~$\Omega$ upon light activation of the photodiode switch. The reason for taking this impedance range is two-fold. First, as the magnitude of the time reflection is proportional to the change in impedance between the two states, we found a 20~$\Omega$ change in impedance would generate a reasonable time reflection, strong enough to be detectable. Second, a State~1 impedance of 60~$\Omega$ minimizes the impedance mismatch of the spatially reflected pulse back into the 50~$\Omega$ measurement equipment. Likewise, a State~2 impedance of 40~$\Omega$ after the time-boundary minimizes the spatial reflections of the reflected and transmitted pulses exiting the DMTL.

Considering the spatial distribution of a 0.5~GHz Gaussian pulse within the DMTL, a total of 16 loading elements with a spacing distance of $d$~=~18.3~mm are chosen to accept the entire pulse completely within the microstrip during the time-boundary event. The final DMTL design parameters are given in Fig.~\ref{fig:equiv_cct}.

Since the experimental results in Section~\ref{sec4} below were taken with an oscilloscope, an example of time-reversal of a two-peak asymmetric pulse reflected and transmitted at a time-boundary is shown in the time-domain at the top of Fig.~\ref{fig:equiv_cct}, referenced at the input voltage $V_\mathrm{in}$ and output voltage $V_\mathrm{out}$. We see within the input pulse (yellow trace), the low-amplitude peak leads the larger peak. After the time-boundary at time $t=t_\mathrm{kick}$, the large amplitude peak leads the smaller one within the reflected wave (red trace) due to time-reversal. This will help with understanding the measured results to follow.

\subsection{A Switchable Capacitive Load with High-Speed Photodiodes}\label{subsec2}

A critical design consideration for our time-boundary experiment is the method of modulating the effective homogeneous medium. The speed at which the time boundary occurs must be within the order of the period of the signal to be scattered \cite{Lustig_2018}. Hence, a Gaussian pulse with a carrier frequency of 0.5~GHz, requires a modulation speed on the order of 1~ns or less. Furthermore, the modulation of each loading element of the periodically-loaded microstrip line must be synchronized to occur at precisely the same time to achieve a truly homogeneous time-switched medium. Finally, the switching element must be readily available and simple to integrate within the circuit. 

To match these requirements, commercially available high-speed PIN photodiodes are used to switch between the two different capacitive states of the microstrip line. Photodiodes offer $ps$-scale rise/fall times, low junction capacitances (${\sim}2~\mathrm{pF}$ or less), and are available in surface mount and fiber-coupled packaging for easy assembly. To activate each photodiode at the same time, a single laser source is triggered, and its output is evenly split to each photodiode using optical fibers for near-perfect (within a few 10s of picoseconds) synchronization.

\begin{figure}[h]
\centering
\includegraphics[width=4.5in]{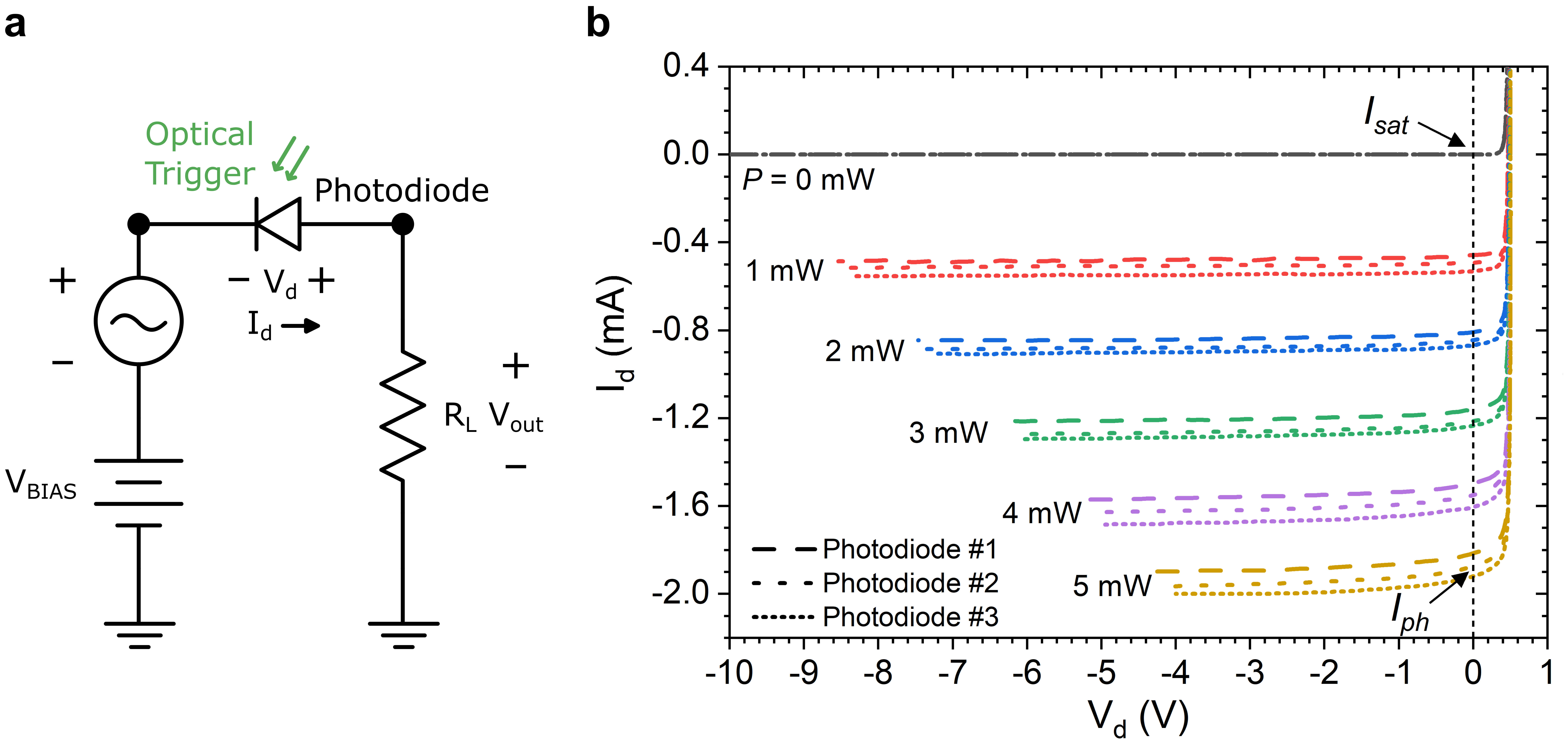}
\caption{\label{fig:pd_iv_curve} \textbf{Equivalent circuit and dc I-V characteristics of a high-speed photodiode operating as a switch. a}, Equivalent circuit diagram of a high-speed photodiode configured to operate as a switch for input signal $V_\mathrm{RF}$. \textbf{b}, the measured dc I-V characteristics of three different photodiodes taken from the same lot, demonstrating the change from the isolating OFF-state ($P$ = 0 mW) to the conductive ON-state (fourth quadrant) at different applied optical powers $P$. The difference between photodiodes for both the OFF-state and ON-state is minimal. For the dc I-V curve measurements, $R_\mathrm{L}=3\,\mathrm{k}\Omega$.}
\end{figure}

The equivalent circuit of a high-speed photodiode switch is shown in Fig.~\ref{fig:pd_iv_curve}~(a). Upon excitation by light, a photocurrent $(I_\mathrm{ph} = R_\mathrm{d}\,P)$ is generated within the photodiode, where $R_\mathrm{d}$ is the photodiode responsivity with units A/W, and $P$ is the applied optical power. Application of this photocurrent transitions the diode from a reverse bias OFF-state (State~1) to a conductive ON-state (State~2). The photodiode operation within a linear regime is simulated using the small-signal approximation \cite{Hernandez_2014}, with the total photodiode current $I_\mathrm{d}$ computed as follows,

\begin{equation}
I_\mathrm{d} = -I_\mathrm{sat}(\mathrm{e}^{qV_\mathrm{d} / nk_\mathrm{B}T} - 1) + I_\mathrm{ph}
\label{eq_3_9}
\end{equation}

\hfill
 
\noindent where $q$ is a unit charge, $n$ is the diode ideality factor, $k_\mathrm{B}$ is the Boltzmann constanst, and $T$ is the temperature.

Sixteen Fermionics FD50S8-F8-FC/APC InGaAs PIN photodiodes are chosen for our design. The datasheet specifications of the photodiodes at $V_\mathrm{bias}$~=~5~V are: rise/fall times of 200~ps max, saturation currents of $I_\mathrm{sat}$~=~0.05~$\pm$~0.03~nA, junction capacitances of $C_{j}$~=~0.33~$\pm$~0.05~pF, and responsivities of $R_\mathrm{d}$~=~0.94~$\pm$~0.04~A/W at 1300~nm. Fig.~\ref{fig:pd_iv_curve}~(b) compares the dc operation of the select group of three photodiodes, depicting their measured dc I-V curves at different optical powers. With increased illuminating power, the magnitude of the diode current in the reverse bias state increases, pushing the operating point of the diodes into the conductive region within the fourth quadrant, showing insignificant deviation in the ON-state regimes of the given photodiodes.

\begin{figure}[t!]
\centering
\includegraphics[width=3in]{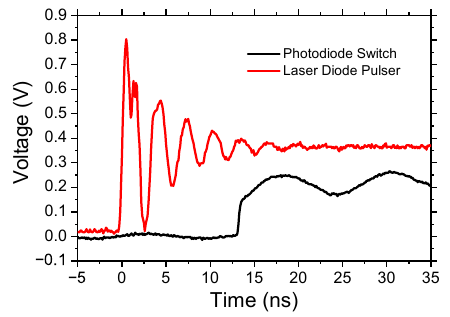}
\caption{\label{fig:pd_speed} \textbf{Measured picosecond rise time of a high-speed photodiode.} Response of the laser diode pulser measured at its monitor pin along with the measured transmission response of the series connected photodiode. A 100~MHz sine wave is sent into the photodiode circuit from a RIGOL DG4102 function generator, and the transmitted signal is captured using a high-speed oscilloscope. The measured 10-90\% rise time (OFF-to-ON) of the 100~MHz signal through the photodiode circuit is 311~ps.}
\end{figure}

To verify the picosecond rise time of the high-speed photodiodes, Fig.~\ref{fig:pd_speed} shows the measured response of both a laser diode pulser and a single photodiode in series. A Highland Technology T165-12 Picosecond Laser Diode Pulser drives a laser diode providing a picosecond optical pulse to the photodiode. In Fig.~\ref{fig:pd_speed}, the measured voltage from an optical waveform monitor pin available on the laser pulser shows a 10-90\% rise time of 283~ps, corresponding to a 10-90\% rise time of a 100 MHz sine wave through the photodiode of 311~ps. The noticable ringing in the laser pulser response is due to EMI artifacts at the monitor pin, which are not present at the laser diode anode.

\section{Experimental Results}\label{sec4}

\begin{figure}[h]
\centering
\includegraphics[width=\textwidth]{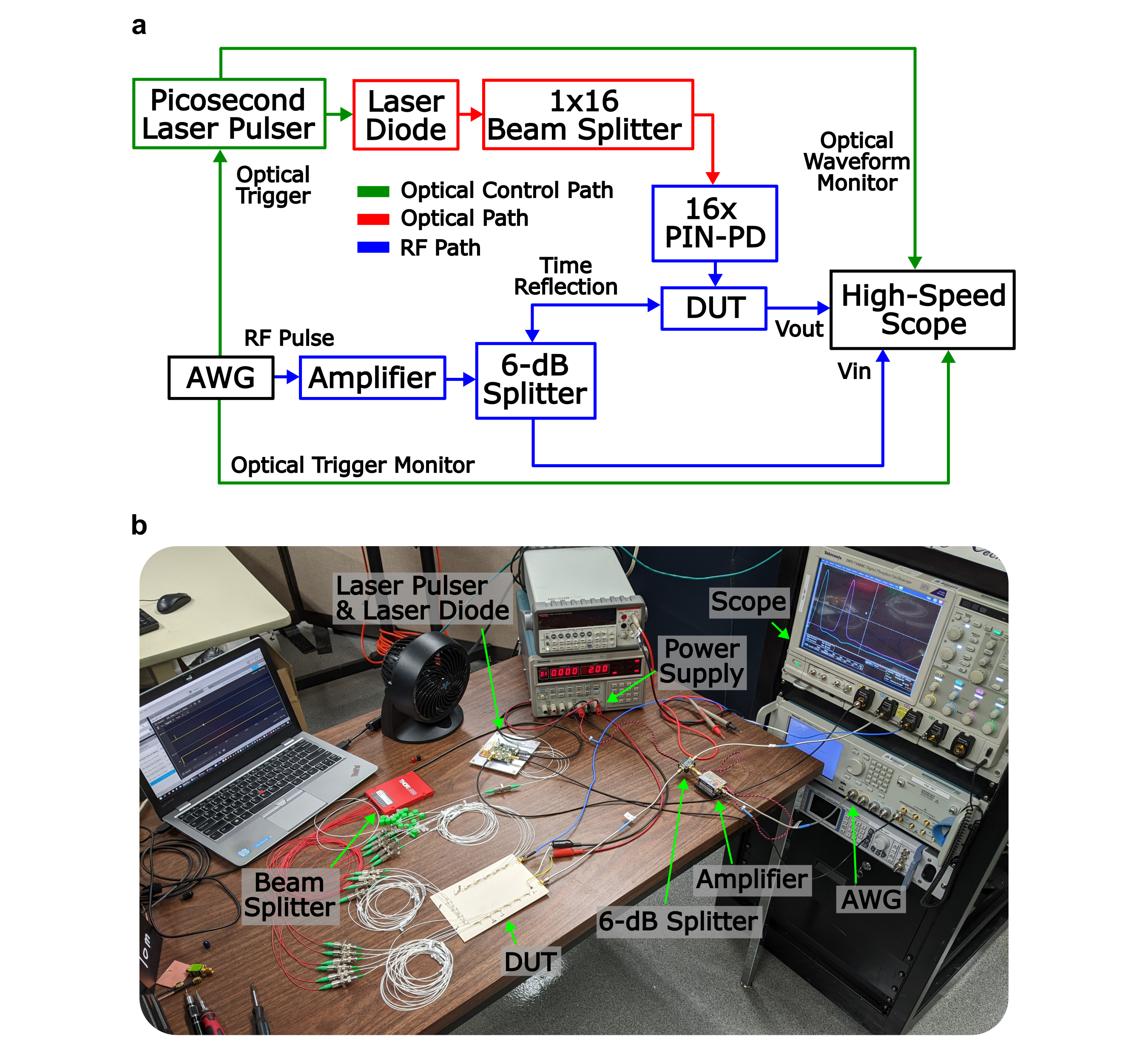}
\caption{\label{fig:exp_setup} \textbf{Experimental setup to demonstrate time reflection. a}, Block diagram. \textbf{b}, Picture of the experimental setup.}
\end{figure}

\begin{figure}[h]
\centering
\includegraphics[width=\textwidth]{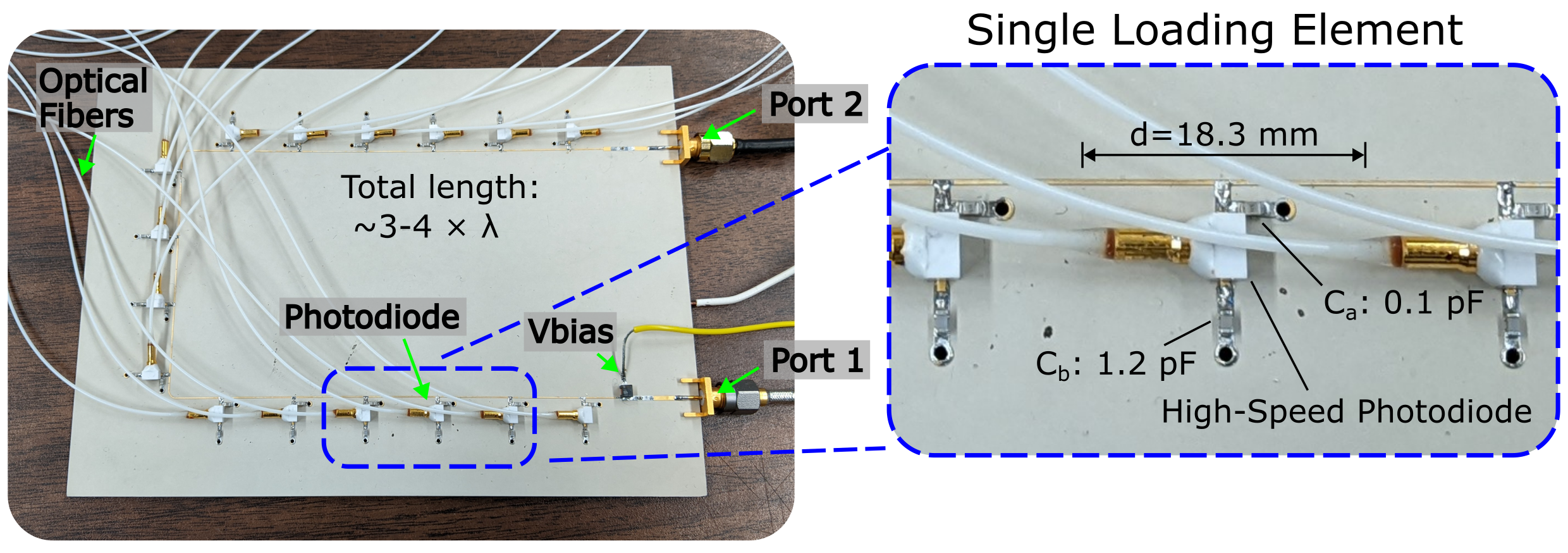}
\caption{\label{fig:dut} \textbf{Fabricated and assembled prototype.} A picture of the fabricated and assembled device under test (DUT) along with a close up of a single loading element with a surface mount high-speed photodiode.}
\end{figure}

Fig.~\ref{fig:exp_setup} shows the block diagram of the time reflection experiment along with a picture of the experimental setup, while Fig.~\ref{fig:dut} depicts a close up of the device under test (DUT) along with a picture of a single loading element. The most critical aspect of this experiment was to synchronize the timing of the high-speed photodiode activation while the RF pulse was within the microstrip TL. To achieve this, a Tektronix AWG700002A Arbitrary Waveform Generator with a sampling rate of 25~GSa/s was used to generate both the optical trigger and RF pulse. This enabled synchronization between the two signals, which could be tuned in 40~ps steps.

A block diagram showing the procedure we used to generate the time reflection is shown in Fig.~\ref{fig:exp_setup}~(a). First, a control signal was sent from an arbitrary waveform generator (AWG) to a Highland Technology T165-12 Picosecond Laser Diode Pulser, which drove an optical pulse with a picosecond rise time from an attached laser diode coupled into an optical fiber. A Thorlabs FPL1053S Fabry-Perot Laser Diode in a butterfly package operating at a wavelength of 1550~nm provides the optical source. This optical pulse is then split evenly into 16 different optical fibers using a Thorlabs TDS1315HA 1x16 beam splitter, with each fiber terminating into a high-speed photodiode periodically loading the microstrip. Considering a tolerance of $\pm$~5~mm for the length of an individual fiber and $n\approx~1.97$ for its refractive index, leads to a $\pm$~33~ps tolerance in synchronized timing between each switch, which is significantly shorter than the switching time and the carrier period of the microwave pulse ($2\,\mathrm{ns}$). Hence, the optical pulses reach all 16 photodiodes at essentially the same time, generating a homogeneous and near-instantaneous change in the microstrip's effective impedance.

Simultaneous to the optical pulse, a microwave pulse is sent into the microstrip from the same AWG carefully synchronized and slightly delayed from the optical control signal such that the microwave pulse is entirely within the microstrip when the optical pulse reaches the photodiodes and activates the time reflection. The microwave pulse then interacts with a time-boundary, with both the transmitted and reflected waves measured with a high-speed oscilloscope. 

A Picosecond Pulse Labs 5330A-140 6-dB power splitter at the input side is used to enable the measurement of the input signal and the time-reflected signal at the scope. The measured insertion loss of the splitter is less than 0.1~dB with a return loss greater than 27~dB up to 1~GHz. It is important to note that both the spatial-reflected and time-reflected signals pass through the power splitter twice before they are recorded by the scope.

An amplifier is used to boost the signal from the AWG due to its maximum output swing of 500~mV\textsubscript{pp}. A Picosecond Pulse Labs 5840A Series amplifier is used, with 41~ps rise time, 80~kHz to 9.3~GHz of bandwidth, 22~dB of gain, +12~dBm saturation, and 5.8~dB noise figure. A specific gain is dialed in to achieve a measured input pulse amplitude of 0.5~V at the scope.

A Tektronix DPO 71604C Digital Phosphor Oscilloscope with 16 GHz bandwidth and 100~GSa/s is used to measure both the input and output signals, the trigger to the laser diode driver, and the monitor pin of the laser diode driver, with an input impedance of 50~$\Omega$ for each channel. The oscilloscope is set to average mode collecting 100 samples for each measurement. 

\begin{figure}[h]
\centering
\includegraphics[width=\textwidth]{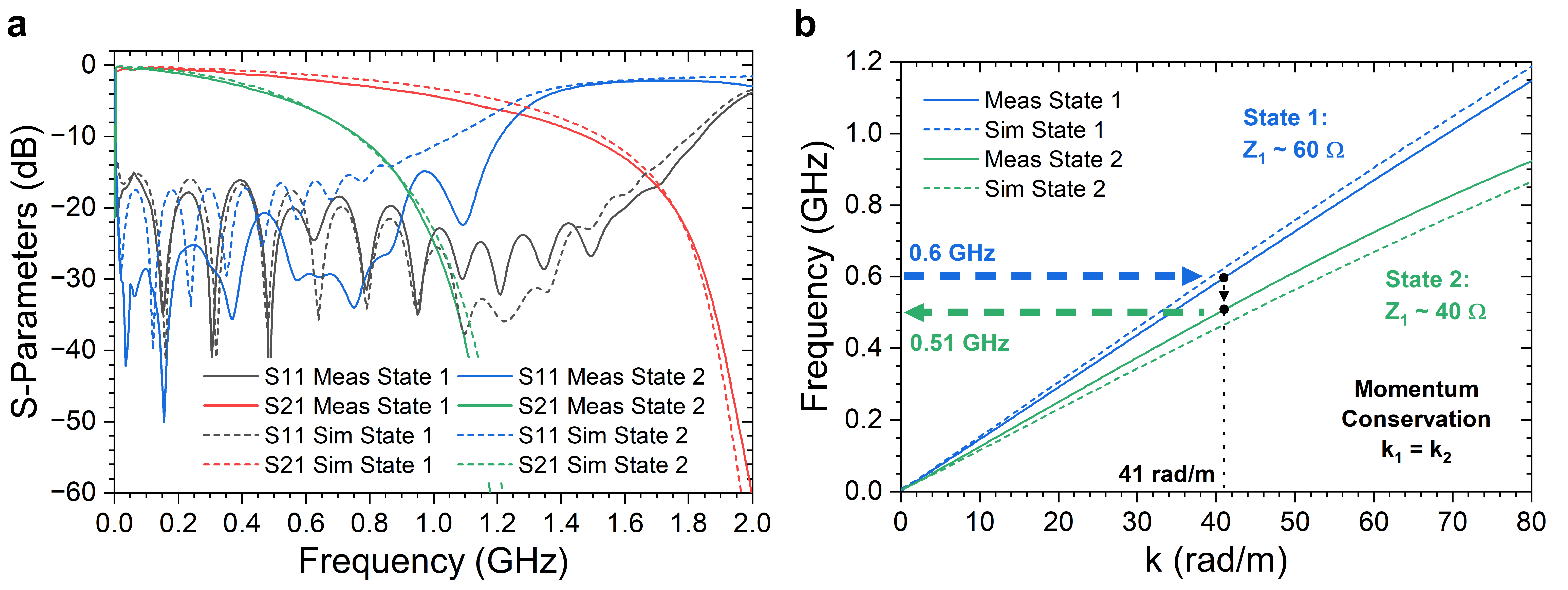}
\caption{\label{fig:freq_resp} \textbf{Measured and simulated frequency response of the periodically loaded microstrip transmission line. a}, Scattering parameters of both State~1 (photodiode switch OFF) and State~2 (photodiode switch ON). \textbf{b}, Corresponding dispersion diagram.}
\end{figure}


\subsection{Frequency Response}

\begin{figure}[h]
\centering
\includegraphics[width=\textwidth]
{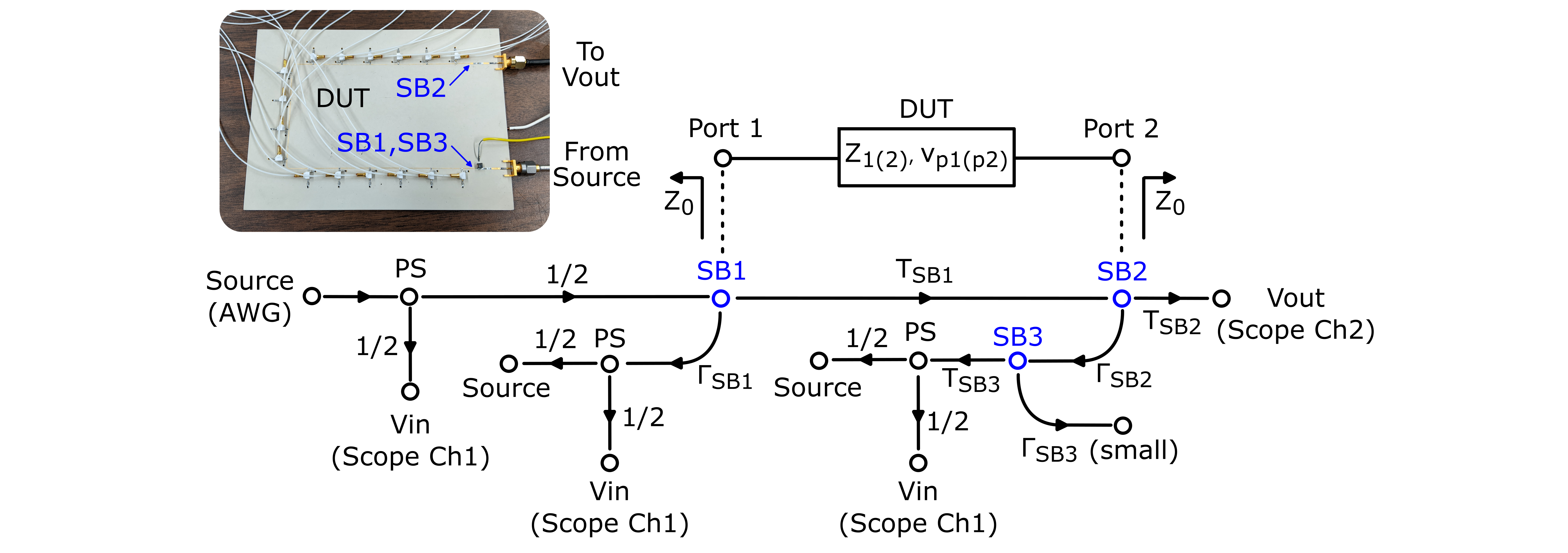}
\caption{\label{fig:signal_flow_no_kick} \textbf{Signal flow graph for the experiment with no time boundary.} The signal enters from the source (AWG), is divided by the 6-dB power splitter (PS), and is then scattered at the first spatial boundary (SB1). The transmitted wave ($T_\mathrm{SB1}$) travels through the periodically-loaded microstrip line (device under test - DUT), and is scattered at the second spatial boundary (SB2) before exiting the DUT ($T_\mathrm{SB2}$) and measured at the output $V_\mathrm{out}$ (channel 2 of the scope). The reflected signal at SB1 ($\Gamma_\mathrm{SB1}$) is again split by the PS and measured at the input $V_\mathrm{in}$ (channel 1 of the scope), with the other half absorbed by the source. The reflected signal at SB2 ($\Gamma_\mathrm{SB2}$) travels back down the DUT, and is again scattered by SB1. The transmitted signal ($T_\mathrm{SB1}$) is split by the PS and measured at $V_\mathrm{in}$. The characteristic impedance $Z_0$ of the system outside the DUT is 50-$\Omega$.}
\end{figure}

Fig.~\ref{fig:freq_resp}~(a) compares the measured and simulated scattering parameters of the periodically loaded microstrip line for the two states, with good agreement. While initially designed for operation up to 1~GHz by limitation of the Bragg frequency cutoff, the measured bandwidth has been limited by the insertion loss, most significantly in State~2. This is due to the $R_\mathrm{on}$ of the photodiode, estimated to be approximately 8~$\Omega$ using the method in \cite{Hernandez_2014}. Losses and parasitics from the discrete components also contribute to the reduced bandwidth. A more optimized photodiode package and periodic loading design using integrated elements could improve the overall bandwidth of the device beyond 1~GHz.

Fig.~\ref{fig:freq_resp}~(b) shows the measured vs. simulated dispersion for the two states, again with good agreement. We use the experimental data to predict the measured frequency translation for a pulse centered at 0.6~GHz. From the conservation of momentum described above in Section~\ref{sec2}, the momentum in State~1 and State~2 are equivalent across the time-boundary, i.e., $k_1$ = $k_2$. Thus, translating the frequency of 0.6~GHz in State~1 to State~2, gives a new frequency of 0.51~GHz, yielding a measured (simulated) frequency translation ratio of $\omega_2/\omega_1$ = 0.85 (0.75), compared to the theoretical frequency translation of 0.67 for an ideal impedance transformation $Z_1 \rightarrow Z_2$ of 60~$\Omega$~$\: \rightarrow \:$~40~$\Omega$. The total length of the periodically-loaded microstrip line is 332.7~mm.

\subsection{Time Reflection of a Single Pulse}

\begin{figure}[t]
\centering
\includegraphics[width=\textwidth]{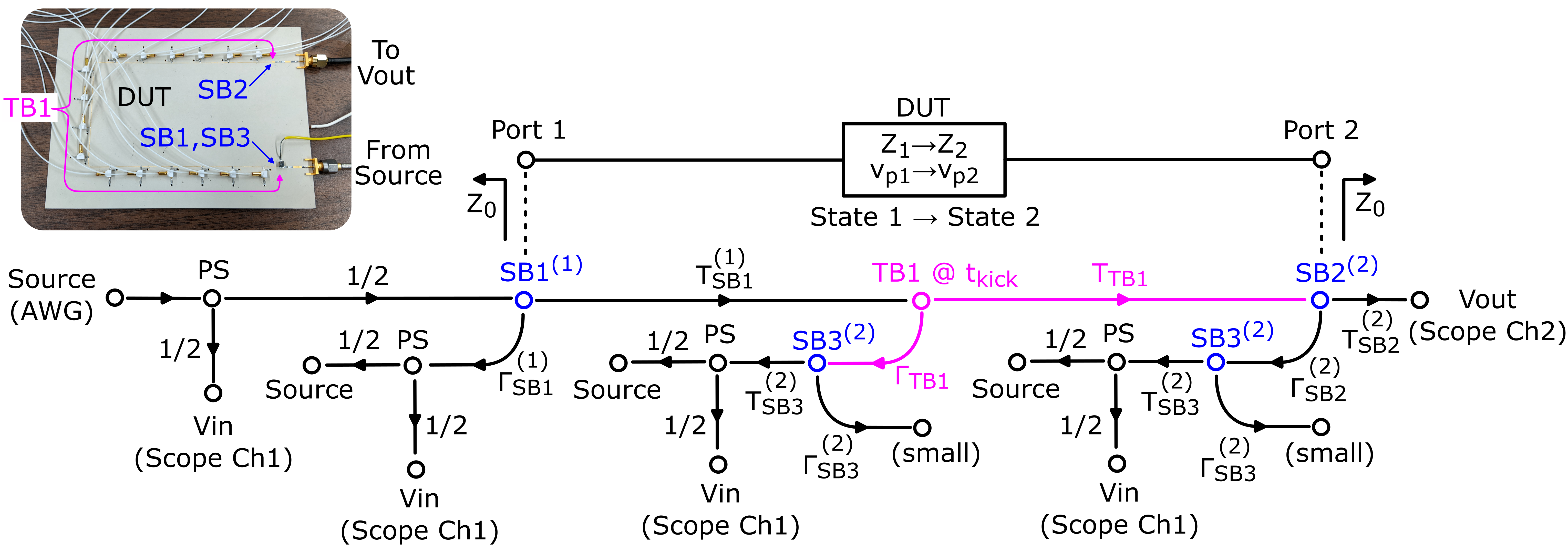}
\caption{\label{fig:signal_flow_with_kick} \textbf{Signal flow graph for the time boundary experiment.} The microstrip line's characteristic impedance and phase velocity is changed from State~1 to State~2 at time $t$~=~$t_\mathrm{kick}$ while the pulse is completely within the DUT. Similar to the case without a time boundary in Fig.~\ref{fig:signal_flow_no_kick}, the signal enters from the source (AWG), is divided by the 6-dB power splitter (PS), and is then scattered at the first spatial boundary SB1\textsuperscript{(1)} in State~1. The transmitted wave travels through the microstrip line in State~1, which is then switched to State~2 while the pulse is completely within the DUT, generating a time boundary TB1 with transmitted and reflected wave coefficients $T_\mathrm{TB1}$ and $\Gamma_\mathrm{TB1}$, respectively. The reflected signal travels back down the microstrip in State~2, and scatters on the first spatial boundary SB1\textsuperscript{(2)}, now in State~2. This transmitted pulse then splits at the PS and enters channel 1 of the scope ($V_\mathrm{in}$), as the measured time-boundary reflection. The time-boundary transmitted signal travels down the microstrip in State~2, and scatters at spatial boundary SB2\textsuperscript{(2)}, also in State~2. This transmitted signal is then measured at $V_\mathrm{out}$ (channel 2 of the scope). The reflected signal travels back down the microstrip line in State~2, scattering at SB1\textsuperscript{(2)}, split at the PS, and measured at $V_\mathrm{in}$.}
\end{figure}

This section presents our simulated and measured results of both the reflected and transmitted signals of a single Gaussian pulse with a full-width half-maximum (FWHM) bandwidth of 0.35~GHz scattered at a time interface. Due to the presence of both spatial boundaries and time boundaries in the experiment, signal flow graphs are provided in Fig.~\ref{fig:signal_flow_no_kick} and Fig.~\ref{fig:signal_flow_with_kick} for the case without and with the time interface, respectively, to provide clarity in the analysis of the simulated and measured results. To extract the time-boundary reflection and transmission coefficients, it is first necessary to extract the spatial-boundary reflection and transmission coefficients in the static case for both State~1 and State~2.

\begin{figure}[h]
\centering
\includegraphics[width=\textwidth]{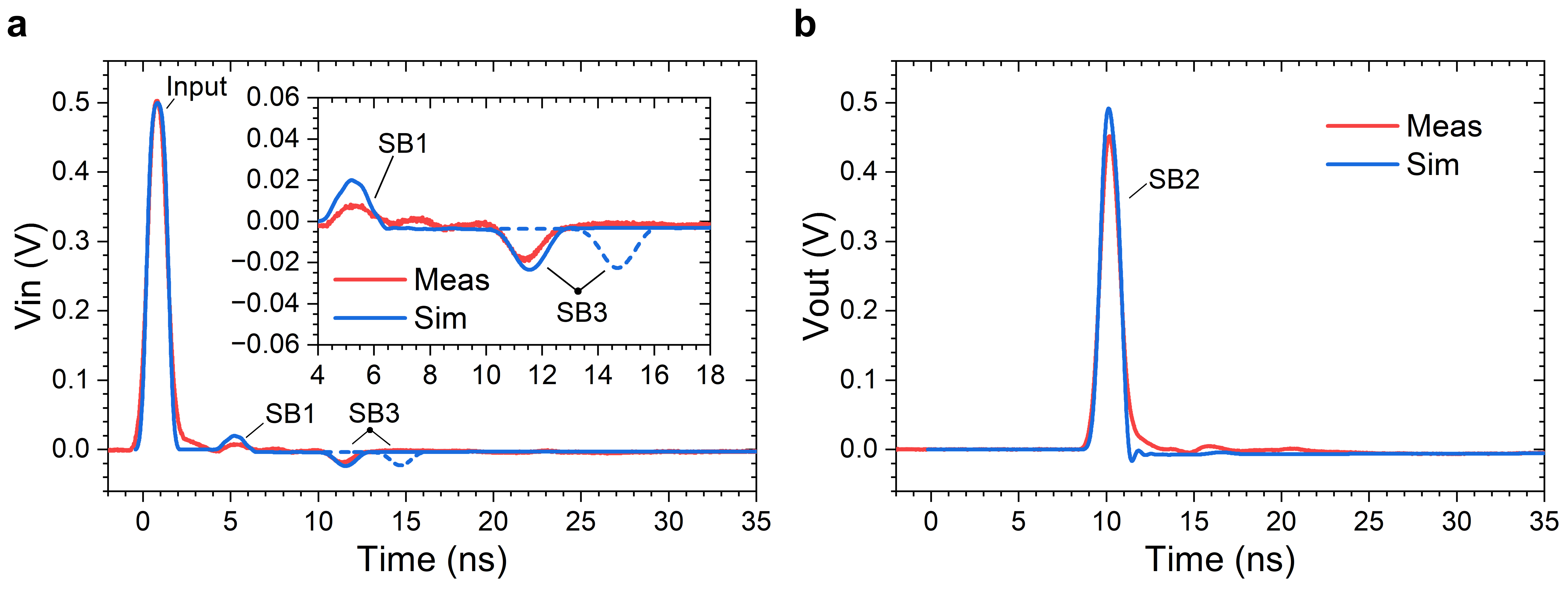}
\caption{\label{fig:no_kick_state_1} \textbf{Measured and simulated response to an input pulse with the microstrip in State~1. a}, The response referenced at the input $V_\mathrm{in}$ to a 0.35~GHz bandwidth Gaussian pulse with the photodiodes in the OFF-state (State~1). \textbf{b}, The output response referenced at $V_\mathrm{out}$ . The dashed blue trace represents the simulated response with 24 loading elements in the model, confirming that the reflected pulses indicated as SB3 in the figure are indeed due to the output spatial boundary SB2 reflected to the input.}
\end{figure}

Based on the signal flow diagram without the time boundary of Fig.~\ref{fig:signal_flow_no_kick}, Fig.~\ref{fig:no_kick_state_1} compares the simulated vs. measured signals when the microstrip is in State~1 for the input $V_\mathrm{in}$ and output $V_\mathrm{out}$, shown in panels (a) and (b) respectively. Using the datasets of these two pulses, we extract the reflection and transmission coefficients at each spatial interface following a procedure detailed in Supplementary Section~2. The simulated and measured results compared to the theoretical values are summarized in Table \ref{tab1}, with good agreement. The discrepancy for the measured $\Gamma_\mathrm{SB1}^{(1)}$ is likely due to the loss in the SMA connector since the spatial boundary is within the microstrip line. This loss is not accounted for in the extraction process.

\begin{table}[h]
\begin{center}
\begin{minipage}{207pt}
\caption{Spatial-Boundary Scattering Coefficients and Characteristic Impedance in State~1}\label{tab1}%
\begin{tabular}{@{}lccccc@{}}
\toprule
  & $\Gamma_\mathrm{SB1}^{(1)}$ & $T_\mathrm{SB1}^{(1)}$ & $\Gamma_\mathrm{SB2}^{(1)}$ & $T_\mathrm{SB2}^{(1)}$ & $Z_{(1)}$ ($\Omega$) \\
\midrule
Theory & 0.091 & 1.091 & -0.091 & 0.909 & 60 \\
Simulation & 0.080 & 1.080 & -0.062 & 0.938 & 56.6 \\
Measurement & 0.030 & 1.030 & -0.075 & 0.925 & 58.1 \\
\botrule
\end{tabular}
\end{minipage}
\end{center}
\end{table}

The reflected pulse from SB2 travels back down the transmission line, and is recorded as a third spatial boundary SB3 in Fig.~\ref{fig:no_kick_state_1}~(a). The scattering coefficients of this third spatial boundary are assumed to be equivalent to the coefficients for SB2. The dashed blue trace in  Fig.~\ref{fig:no_kick_state_1}~(a) represents the simulated model with 24 loading elements, indicating the pulse at SB3 is indeed due to the reflection off SB2 and transmitted through SB3 back to the input.

\begin{figure}[h]
\centering
\includegraphics[width=\textwidth]{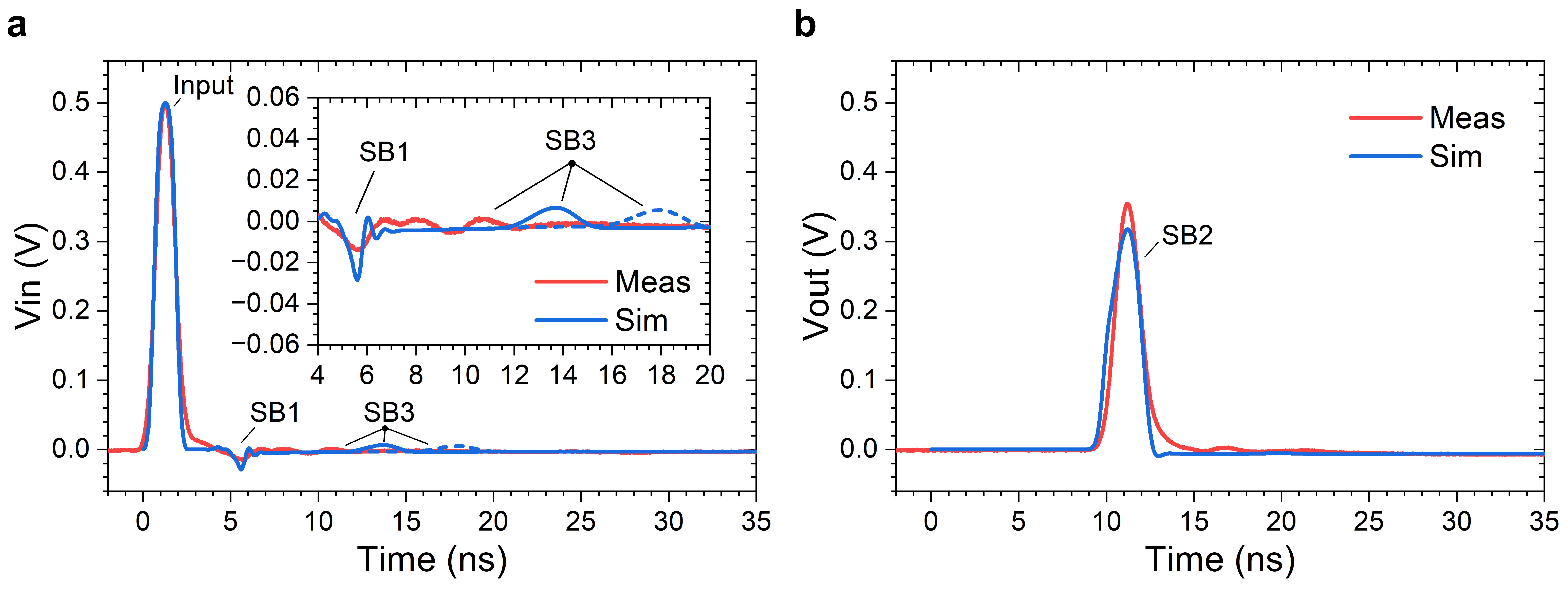}
\caption{\label{fig:no_kick_state_2} \textbf{Measured and simulated response to an input pulse with the microstrip in State~2. a}, The response referenced at the input $V_\mathrm{in}$ to a 0.35~GHz bandwidth Gaussian pulse with the photodiodes in the ON-state (State~2). \textbf{b}, The output response referenced at $V_\mathrm{out}$. The dashed blue trace again represents the simulated response with 24 loading elements in the ADS model.}
\end{figure}

Figs.~\ref{fig:no_kick_state_2}~(a) and (b) compare the simulated vs. measured results for the static case when the microstrip line is in State~2, with the ON-state photodiodes. The optical pulse is turned on for approximately 52~ns for the measured results to capture all dynamic events. The response at the input $V_\mathrm{in}$ is plotted in (a), showing the reflected pulse from the first spatial boundary SB1. The response at the output $V_\mathrm{out}$ is plotted in (b), showing the transmitted pulse through the second spatial boundary SB2. The method detailed in Supplementary Section~2 is again used to extract the reflected and transmitted coefficients, summarized in Table \ref{tab2} and compared to the theoretical values with good agreement. The discrepancy in measured $\Gamma_\mathrm{SB1}^{(2)}$ is again likely due to the SMA connector loss, similar to the State~1 analysis above. The dashed blue trace in (a) represents the simulated response with 24 elements, again demonstrating the pulse at SB3 is due to the reflection off the output spatial boundary SB2 and captured at the input, where it is delayed with more loading elements. The simulated and measured pulses at 13.6~ns and 10.7~ns, respectively, labeled SB3, are due to this reflection. 

\begin{table}[h]
\begin{center}
\begin{minipage}{207pt}
\caption{Spatial-Boundary Scattering Coefficients and Characteristic Impedance in State~2}\label{tab2}%
\begin{tabular}{@{}lccccc@{}}
\toprule
  & $\Gamma_\mathrm{SB1}^{(2)}$ & $T_\mathrm{SB1}^{(2)}$ & $\Gamma_\mathrm{SB2}^{(2)}$ & $T_\mathrm{SB2}^{(2)}$ & $Z_{2}$ ($\Omega$) \\
\midrule
Theory & -0.111 & 0.888 & 0.111 & 1.111 & 40 \\
Simulation & -0.114 & 0.886 & 0.128 & 1.128 & 38.6 \\
Measurement & -0.058 & 0.942 & 0.061 & 1.061 & 44.2 \\
\botrule
\end{tabular}
\end{minipage}
\end{center}
\end{table}

The discrepancy of 2.9~ns between the simulated and measured pulses at SB3 is likely due to differences in the simulated and measured phase velocity of State~2. The phase velocities of the two states are extracted from the dispersion diagram in Fig.~\ref{fig:freq_resp}~(b). For State~1, the simulated and measured phase velocities $\nu_\mathrm{p1}$ are 9.6~m/s and 9.2~m/s, respectively. For State~2, the simulated and measured velocities $\nu_\mathrm{p2}$ are 7.2$\times$10\textsuperscript{7}~m/s and 7.9$\times$10\textsuperscript{7}~m/s, respectively. The slower speed and larger discrepancy in the simulated and measured phase velocities of State~2 corroborates this discrepancy.

\begin{figure}[h]
\centering
\includegraphics[width=\textwidth]{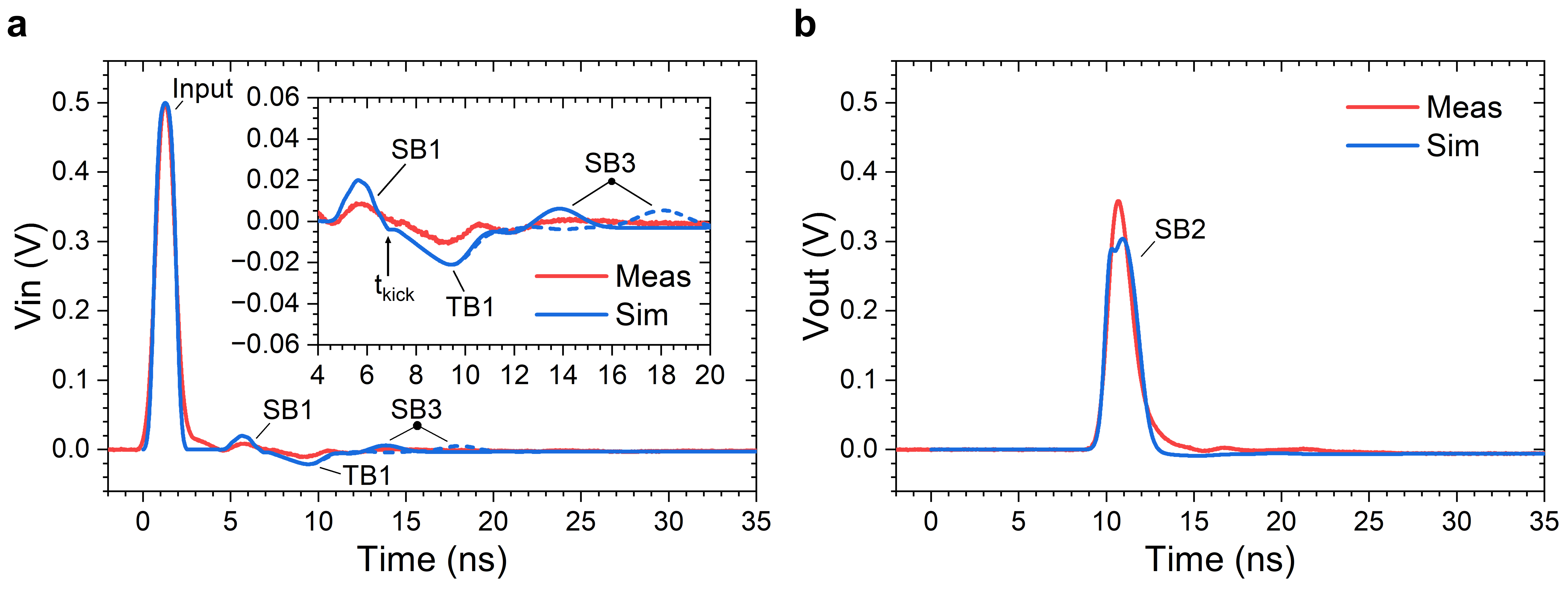}
\caption{\label{fig:with_kick_single_pulse} \textbf{Measured and simulated response of the time-boundary experiment. a}, The response to a 0.35~GHz bandwidth Gaussian pulse referenced at the input $V_\mathrm{in}$ with a time boundary at time $t_\mathrm{kick}$~=~6.85~ns. \textbf{b}, Output response referenced at $V_\mathrm{out}$. The dashed blue trace again represents the simulated response with 24 loading elements in the ADS model confirming that the reflected pulse indicated as SB3 in the figure is due to the output spatial boundary SB2 reflected to the input.}
\end{figure}

With the spatial coefficients known, the time-boundary coefficients are now extracted. Referring to the signal flow graph with the time boundary in Fig.~\ref{fig:signal_flow_with_kick}, the simulated and measured results are plotted in Fig.~\ref{fig:with_kick_single_pulse}~(a) and (b) referenced at the input voltage $V_\mathrm{in}$ and output voltage $V_\mathrm{out}$, respectively. The time boundary is activated at $t=t_\mathrm{kick}$, when the pulse has fully entered but has yet to exit the microstrip trace. As mentioned previously in Section~\ref{sec3}, the total length of the microstrip line in State~1 was chosen to be just long enough to contain the spatial spread of a single pulse, thus the pulse has travelled approximately halfway through the line before the time-boundary is activated.

Referring to Fig.~\ref{fig:with_kick_single_pulse}~(a), the reflected pulse due to the first spatial boundary at SB1 is clearly visible with a positive amplitude due to the microstrip line being in State~1. The reflected pulse due to the output spatial boundary SB2 and indicated by SB3 at the input is also clearly visible with positive amplitude due to the microstrip line now being in State~2. However, we now see a pulse with negative amplitude between these two points, indicated by TB1. This pulse is due to the time interface event triggered at time $t_\mathrm{kick}$~=~6.85~ns. The time-boundary reflection and transmission coefficients are extracted using the method described in Supplementary Section~2, and are plotted along with the theoretical values in Table~\ref{tab3}, showing good agreement. 

\begin{table}[h]
\begin{center}
\begin{minipage}{107pt}
\caption{Time-Boundary Scattering Coefficients}\label{tab3}%
\begin{tabular}{@{}lcc@{}}
\toprule
  & $\Gamma_\mathrm{TB1}$ & $T_\mathrm{TB1}$ \\
\midrule
Theory & -0.111 & 0.556 \\
Simulation & -0.084 & 0.604 \\
Measurement & -0.048 & 0.780 \\
\botrule
\end{tabular}
\end{minipage}
\end{center}
\end{table}

The slight discrepancy in the measured $\Gamma_\mathrm{TB1}$ is likely due to losses and dispersion in the measured prototype unaccounted for in the extraction process, as well as the finite switching speed of the photodiodes. While their switching speed was confirmed to be 311~ps, as discussed above in Section~\ref{sec3}, the period of a 350~MHz signal is ${\sim}$2.9~ns. While the switching time is faster than the period of the wave traveling through the medium, a requirement for time-crystals \cite{Lustig_2018}, a faster switching time would lead to a more abrupt time interface. This would lead to an increased reflection amplitude similar to an equivalent abrupt spatial interface, as explored in \cite{Moussa_2023}.

\begin{figure}[h]
\centering
\includegraphics[width=\textwidth]{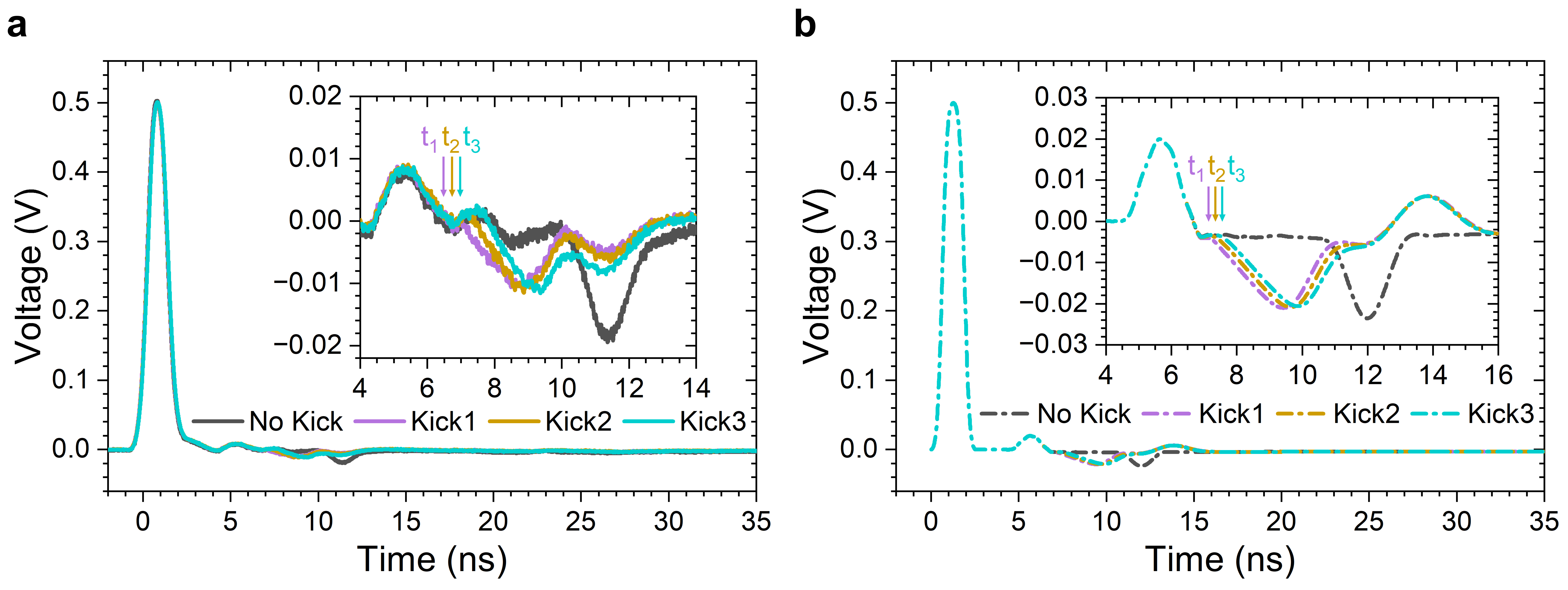}
\caption{\label{fig:single_pulse_time_shift} \textbf{Measured and simulated time reflection of a single Gaussian pulse while adjusting the triggering of the time interface. a}, Measured response referenced at the input $V_\mathrm{in}$ while adjusting the triggering of the time interface with three different steps at $t_1$~=~6.47~ns, $t_2$~=~6.75~ns, and $t_3$~=~6.97~ns. \textbf{b}, Simulated response corresponding well to measurement.}
\end{figure}

To further demonstrate the pulse reflection and transmission indicated by TB1 is due to a time boundary, Fig.~\ref{fig:single_pulse_time_shift}~(a) and (b) plot the measured and simulated response, respectively, of a single Gaussian pulse in which the triggering of the time-interface $t_\mathrm{kick}$ is varied. In Fig.~\ref{fig:single_pulse_time_shift}~(a), we see that by delaying the triggering of the kick from $t_1$ to $t_3$, the time-reflected pulse at TB1 is delayed accordingly. In Fig.~\ref{fig:single_pulse_time_shift}~(b), we see a similar response in the simulated model, thus confirming this pulse is generated by the time-boundary event.

\subsection{Time Reflection of a Two-Peak Asymmetric Pulse}\label{subsec4.3}

\begin{figure}[h]
\centering
\includegraphics[width=\textwidth]{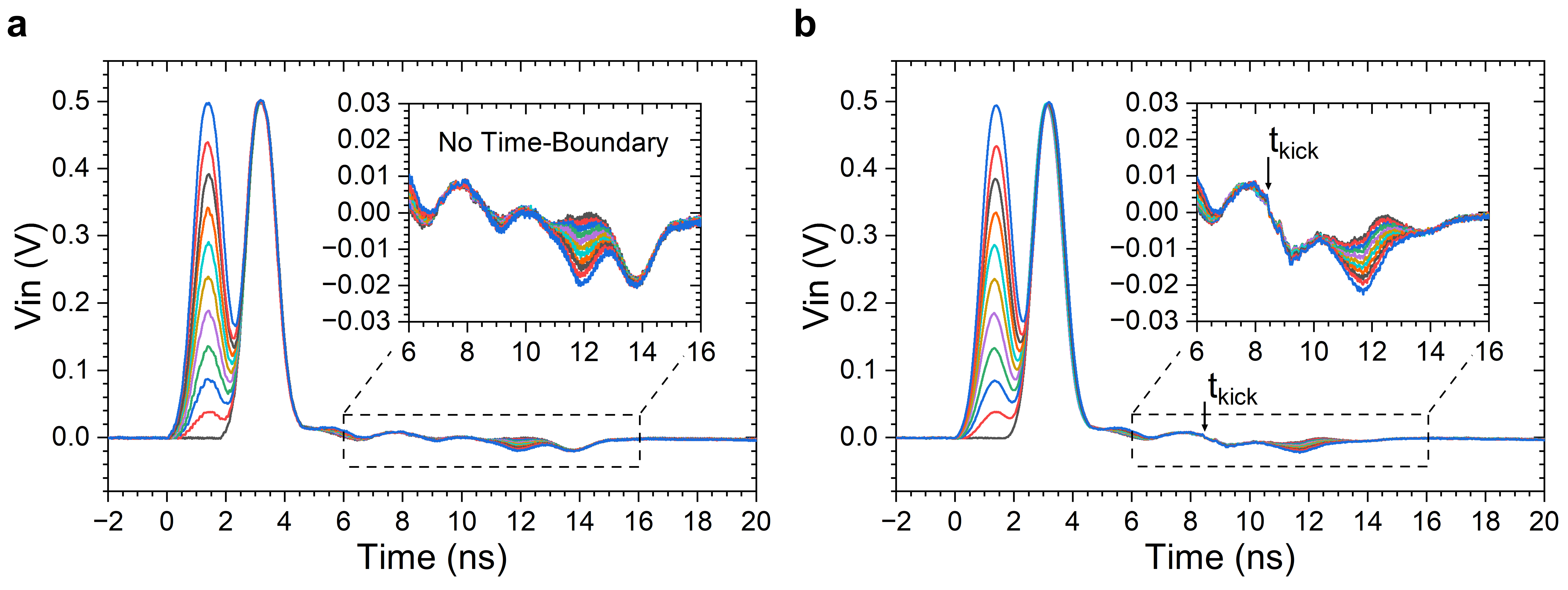}
\caption{\label{fig:double_pulse} \textbf{Measured results of two sequential pulses fed into the microstrip line in which the amplitude of the first pulse is varied. a}, Demonstration of a spatial reflection without a time interface. \textbf{b} Demonstration of a time reflection with the time interface occurring at $t_\mathrm{kick}$~=~8.4~ns.}
\end{figure}


Fig.~\ref{fig:double_pulse} illustrates the experimental demonstration of the time-reversal property of reflected waves at a time-boundary discussed in Section~\ref{sec2} and Section~\ref{sec3} above. The figure shows the measured input voltage versus time of a two-peak asymmetric pulse fed into the microstrip and the corresponding reflection signals without (a) and with the time interface (b). By varying the amplitude of the first pulse, it is clear that (a) demonstrates a spatial reflection, i.e., the leading pulse is the first to reflect due to the impedance mismatch at the output and return to the input. In contrast, by applying a time-boundary at time $t_\mathrm{kick}$~=~8.4~ns, the time at which the pulse has fully entered the microstrip line but has yet to exit, (b) demonstrates a time reflection. Here, the leading input peak becomes the lagging peak due to the time reversal property of the time interface, a homogeneous and near-instantaneous change in the effective characteristic impedance of the microstrip. Likewise, after the time-boundary, the lagging input pulse becomes the leading peak of the reflected wave. The FWHM bandwidth of each peak is approximately 0.4~GHz.

\subsection{Frequency Translation Demonstration}\label{subsec3}

\begin{figure}[t]
\centering
\includegraphics[width=\textwidth]{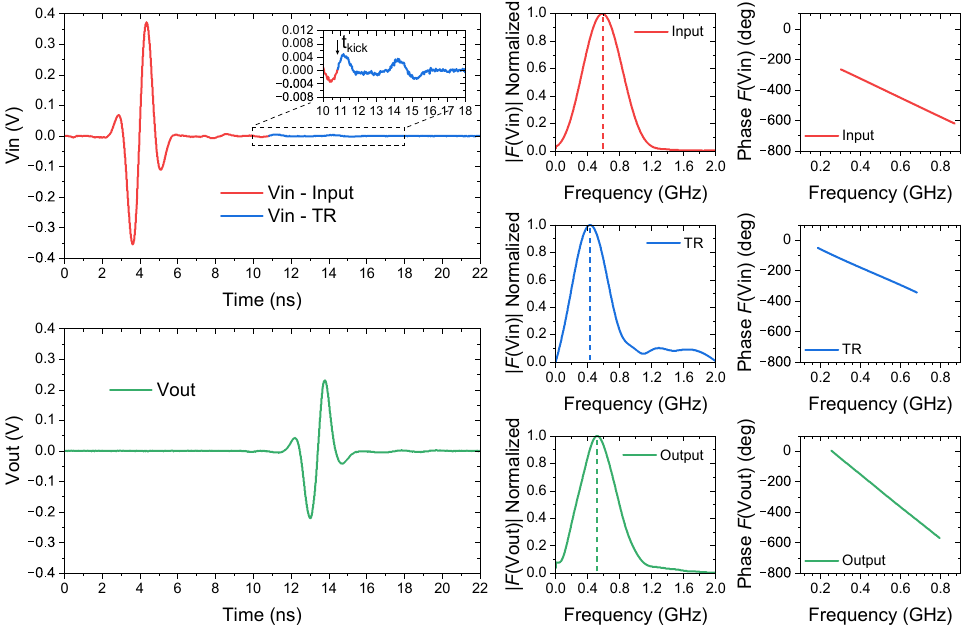}
\caption{\label{fig:translation} \textbf{Measured results of a single carrier Gaussian input pulse scattered at a time interface.} The input pulse has a carrier frequency of 0.59~GHz incident at a time-boundary triggered at $t$~=~$t_\mathrm{kick}$, with the corresponding time-reflected and transmitted pulses demonstrating frequency translation.}
\end{figure}

To demonstrate frequency translation at a time-boundary as discussed in Section~\ref{sec2} with our prototype, the time reflection and transmission of an input Gaussian pulse with a single carrier frequency was measured. Time-gated Fourier transforms are applied independently over the input (red trace), reflected (blue trace), and transmitted (green trace) to extract their spectral and phase content, with the results plotted in Fig.~\ref{fig:translation}. The spectral phase versus frequency are plotted across the FWHM of each pulse.

The input pulse in the frequency domain is centered at 0.59~GHz. The time-reflected pulse is centered at 0.43~GHz, and the transmitted pulse is centered at 0.51~GHz. This is conclusive evidence of frequency translation in our prototype, and also demonstrates the reflection of an electromagnetic pulse with the highest-to-date frequency from an optically-controlled ultra-fast time interface. The frequency translation corresponds to an $\omega_2/\omega_1$ ratio of 0.74 and 0.88 for the time-reflected and transmitted waves, respectively, compared to the expected 0.67 from the theory presented in Section~\ref{sec2} above. The discrepancy between measurement and theory is thought to be due to the finite length of the transmission line. Part of the pulse has begun to exit the microstrip line just before the photodiodes have completely turned on, triggering the time interface.

Regarding the spectral phase plots in Fig.~\ref{fig:translation}, the input pulse has a phase of -81.7 degrees at 0.59~GHz, the time-reflected pulse has a phase of -196.6~degrees at 0.43~GHz, and the transmitted pulse has a phase of -273.4~degrees at 0.51~GHz. Extrapolating the curves back to dc, the time-reflected signal exhibits a close to 180-degree shift compared to the input and time-refracted signals, corresponding to the theory prediction \cite{Moussa_2023}. The slope of the time-reflected signal is also shallower than the input signal, indicating negative phase accumulation (counter-clockwise rotating phase vector) of the signal after the time boundary. This interesting effect will be explored in the next section. 

\subsection{Time Reversal and Phase Conjugation Demonstration}\label{subsec4}

\begin{figure}[h]
\centering
\includegraphics[width=\textwidth]{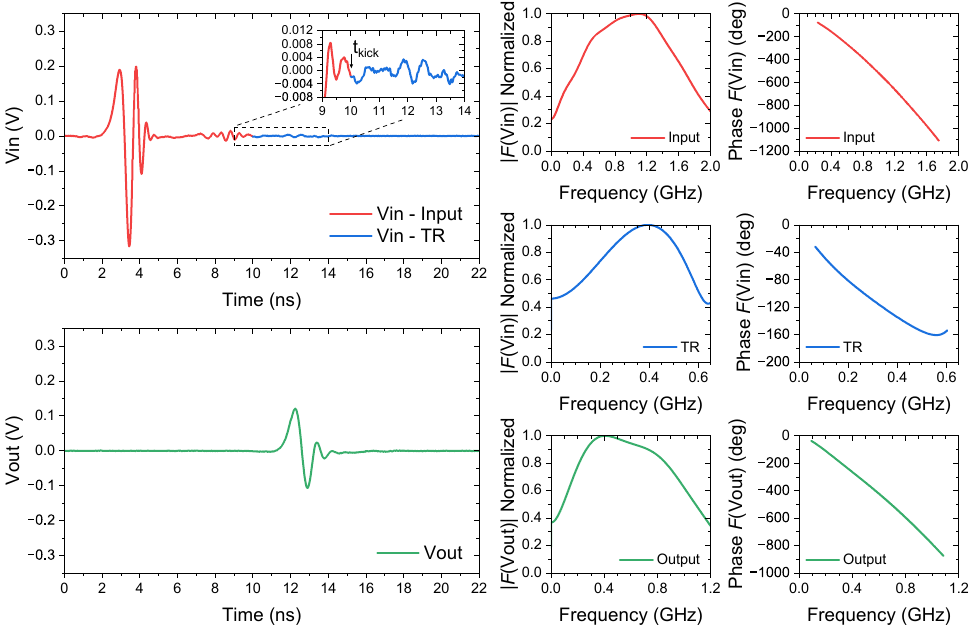}
\caption{\label{fig:reversal} Measured results of an input chirped pulse with a Gaussian envelope and nonlinear phase and the corresponding time-reflected and transmitted pulses demonstrating both frequency translation and time reversal.}
\end{figure}

This section provides experimental evidence of the phase-conjugation nature of time-reflected waves \cite{Lerosey_2004,Biancalana_2007,Bacot_2016,Lustig_2018}. From the voltage relationship (\ref{eq_a19}) derived in Supplementary Section~1, and considering the transmitted and reflected waves have the same angular frequency magnitude but are 180 degrees out of phase, i.e., $\omega_{2-T} = (-)\omega_{2-R}$, the transmitted and time-reflected waves can be split up accordingly \cite{Mendonca_2002}:

\begin{equation}
V_\mathrm{2-T}=V_{01}Te^{j(\omega_\mathrm{2-T}t-kx)}=V_{01}T[cos(\omega_\mathrm{2-T}t-kx)+jsin(\omega_\mathrm{2-T}t-kx)]\label{eq_30}
\end{equation}

\begin{equation}
V_\mathrm{2-R}=V_{01}Re^{-j(\omega_\mathrm{2-R}t+kx)}=V_{01}R[cos(\omega_\mathrm{2-R}t+kx)-jsin(\omega_\mathrm{2-R}t+kx)]. \label{eq_31}
\end{equation}

\hfill

\noindent Here $V_\mathrm{2-T}$ and $V_\mathrm{2-R}$ represent the transmitted and time-reflected waves, respectively, while $\omega_\mathrm{2-T}$ and $\omega_\mathrm{2-R}$ represent the transmitted and reflected waves' angular frequency, respectively. These two waves have also been split into their quadrature components to highlight the phase rotation for each. 

We see that the transmitted wave in (\ref{eq_30}) rotates counter-clockwise, such that the wave incurs positive phase as it propagates in the positive $x$ direction. However, the phase rotation for the time-reflected wave in (\ref{eq_31}) is clockwise, such that the wave incurs negative phase as it propagates in the negative $x$ direction. Thus, the time-reflected wave is essentially a time-reversed wave, where any phase incurred in the wave before the time-reflection can be reversed in the time-reflected wave after the time-reflection. 

To measure this time-reversal property, a chirped pulse with a Gaussian envelope and nonlinear phase was sent into our microstrip prototype and time reflected. The details of the input chirped pulse can be found in Supplementary Section~3. The experimental results are plotted in Fig. \ref{fig:reversal}. Again, frequency translation of the input pulse is observed, as expected, and the phase versus frequency of both the input and transmitted pulses have a negative slope. However, the phase of the time-reflected pulse has a positive slope, indicating phase-conjugation of the reflected wave.





\section{Conclusion}\label{sec6}

We experimentally demonstrate the reflection and transmission of electromagnetic waves from a broadband microwave pulse with the highest-to-date carrier frequency of 0.59~GHz due to an optically-controlled ultra-fast time interface. The design and implementation of this time-boundary is enabled by a reconfigurable periodically-loaded microstrip transmission line capable of achieving near-instantaneous change in its effective impedance between two states using commercially-available high-speed photodiodes with picosecond switching rates. The prototype is optimized to incorporate advanced yet realistic and accessible design solutions with practical advantages for real-world applications in future microwave systems. Excellent agreement between theory, simulation, and measurement is shown, validating the unique properties of time-reflected waves at microwave frequencies, including the time-reversal, frequency translation, and phase-conjugate nature of time-reflected waves.

\backmatter




\bmhead{Acknowledgments}

This work is supported by the AFOSR under FY21-MURI award number FA9550-21-1-0299 (Dr. Arje Nachman, Program Manager).





%

\newpage

\section*{Supplementary Information}

\renewcommand\thefigure{S\arabic{figure}}
\setcounter{figure}{0} 

\renewcommand{\theequation}{S\arabic{equation}}
\setcounter{equation}{0} 





\noindent \textbf{S1. Derivation of the Reflection and Transmission Coefficients of a Time Boundary}

\begin{figure}[ht]
\centering
\includegraphics[width=3in]{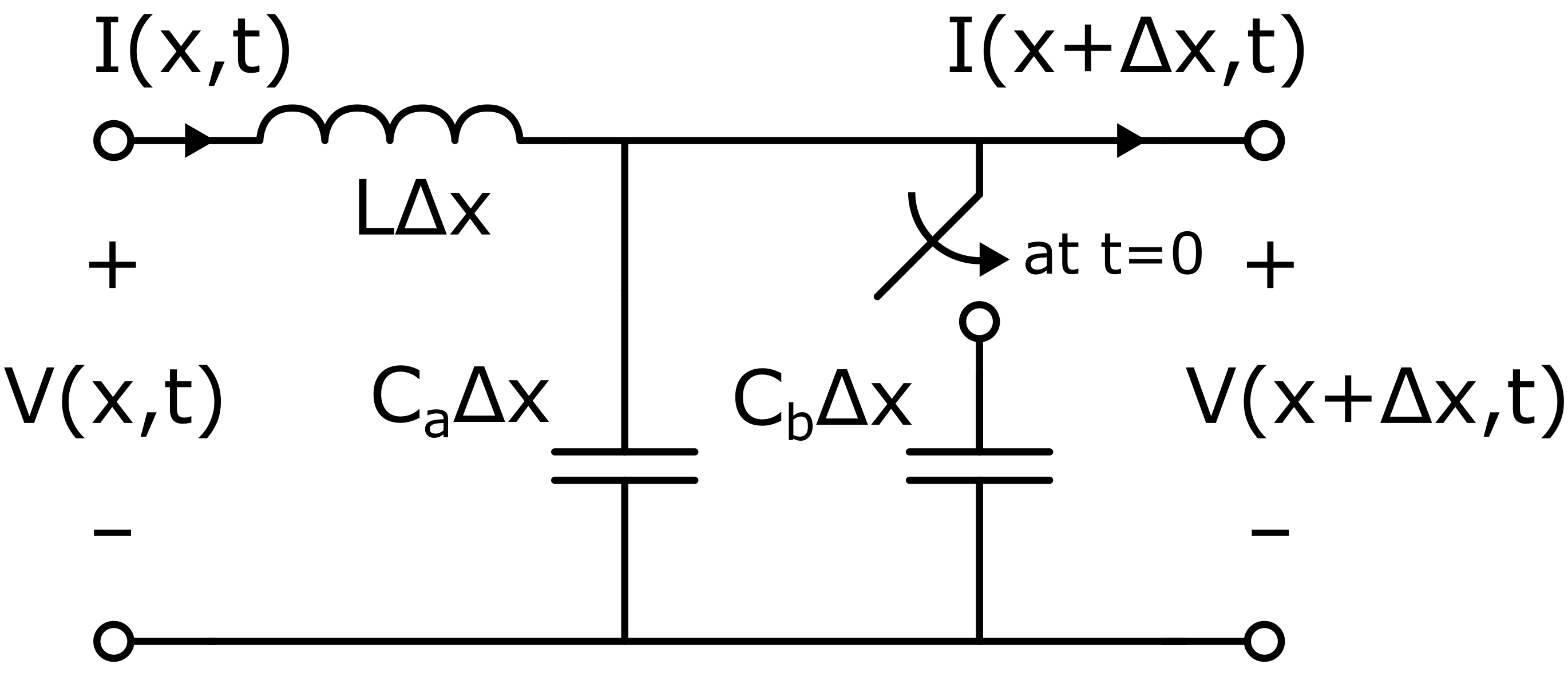}
\caption{\label{fig:distributed_circuit_model} Distributed circuit diagram of a microwave transmission line for the case when an additional capacitive loading is switched on at time $t$ = 0.}
\end{figure}

\indent Fig.~\ref{fig:distributed_circuit_model} shows the distributed circuit diagram of a microwave transmission line (TL) used to derive the reflection $R$ and transmission $T$ coefficients due to the scattering of a microwave signal at a time interface. This time interface is generated by switching on a second shunt capacitive loading within the TL, increasing the effective permittivity and decreasing the characteristic impedance of the line. Before the switching event, only the shunt capacitor $C_\mathrm{a}$ loads the TL, for a total capacitance of $C_{1}=C_\mathrm{a}$. At time $t=0$, the switch is thrown, and the additional shunt capacitance $C_\mathrm{b}$ is added to the circuit, for a total capacitance of $C_{2}=C_\mathrm{a}+C_\mathrm{b}$. Referencing the treatment in \cite{Moussa_2023}, the corresponding telegraphers equations at time $t>0$ are derived beginning with \cite{Pozar_2012}:

\begin{equation}
v(x,t) = L\Delta x\frac{\partial i(x,t)}{\partial t}+v(x+\Delta x,t) \label{eq_a1}
\end{equation}

\begin{equation}
i(x,t) = C_\mathrm{a}\Delta x\frac{\partial v(x+\Delta x,t)}{\partial t} + C_\mathrm{b}\Delta x\frac{\partial v(x+\Delta x,t)}{\partial t} + i(x+\Delta x,t) \label{eq_a2}
\end{equation}

\hfill

Dividing both equations by $\Delta x$ and taking the limit $\Delta x\to0$ yields the telegrapher's equations:

\begin{equation}
\frac{\partial v(x,t)}{\partial t} = -L\frac{\partial i(x,t)}{\partial t} \label{eq_a3}
\end{equation}

\begin{equation}
\frac{\partial i(x,t)}{\partial t} = -C_\mathrm{a}\frac{\partial v(x,t)}{\partial t} - C_\mathrm{b}\frac{\partial v(x,t)}{\partial t} \label{eq_a4}
\end{equation}

\hfill

Multiplying (\ref{eq_a3}) by $\frac{\partial}{\partial x}$ and (\ref{eq_a4}) by $\frac{\partial}{\partial t}$, and then plugging (\ref{eq_a4}) into (\ref{eq_a3}) yields the equation:

\begin{equation}
\frac{\partial^2 v(x,t)}{\partial t^2} = LC_\mathrm{a}\frac{\partial^2 v(x,t)}{\partial t^2}+LC_\mathrm{b}\frac{\partial^2 v(x,t)}{\partial t^2} \label{eq_a5}
\end{equation}

\hfill

A Laplace transform is now applied on the voltage $v(x,t)$ according to $\mathcal{L}\{v(x,t)\}=\int_{0}^{\infty} v(x,t)\mathrm{e}^{-st}dt = V(x,s)$ and $\mathcal{L}\{\frac{\partial^2 v(x,t)}{\partial t^2}\} = s^2V(x,s) - sv(x,0^{-}) - \frac{\partial v(x,0^{-})}{\partial t}$ \cite{Nilsson_2012}, and assuming a time-harmonic $\mathrm{e}^{j\omega t}$ variation of the voltage and currents, the above equation becomes:

\begin{equation}
\begin{split}\label{eq_a6}
\frac{\partial^2 v(x,t)}{\partial t^2} &= LC_\mathrm{a}\{ s^2V(x,s) - sv(x,0^{-}) \}\\ &+ LC_\mathrm{b}\{ s^2V(x,s) - sv(x,0^{-}) - j\omega_{1}v(x,0^{-}) \} 
\end{split}
\end{equation}

\hfill

Since at $t<0$ the voltage drop across $C_\mathrm{b}$ is equal to zero, and recognizing $C_{2}=C_\mathrm{a}+C_\mathrm{b}$ and $C_{1}=C_\mathrm{a}$, the above equation can be simplified to the following wave equation for the voltage:

\begin{equation}
\left(\frac{\partial^2}{\partial x^2} - LC_{2}s^2\right)V(x,s) = -LC_{1}(s + j\omega_{1})v(x,0^{-})\label{eq_a7}
\end{equation}

\hfill

\noindent where $\omega_{1}$ is the angular frequency of the signal before the time boundary. Assuming a harmonic variation in space  $\mathrm{e}^{-jkx}$ gives the voltage,

\begin{equation}
V(x,s) = \frac{C_{1}(s + j\omega_{1})}{C_{2}(\omega_{2}^2 + s^2)}v(x,0^{-})\label{eq_a8}
\end{equation}

\hfill

\noindent where $\omega_{2}$ is the angular frequency of the signal after the time boundary, due to frequency translation. Since the TL is effectively a homogeneous medium and there is no variation in $x$ across the time barrier, the conservation of momentum between the two states yields,

\begin{equation}
k = \omega_{1}\sqrt{LC_{1}} = \omega_{2}\sqrt{LC_{2}} \label{eq_a9}
\end{equation}

\begin{equation}
\omega_{2} = \sqrt{\frac{C_{1}}{C_{2}}}\omega_{1} = \frac{Z_{2}}{Z_{1}}\omega_{1} \label{eq_a10}
\end{equation}

\hfill

\noindent where $Z_{1}$ and $Z_{2}$ are the characteristic impedance of the TL before and after the time boundary, respectively. Equation (\ref{eq_a10}) predicts the frequency translation across the time boundary as a function of the ratio of the two capacitive loading states, or the impedances of the two different states. The above assumes a time-harmonic field, where the second order time derivative on the right side of (\ref{eq_a5}) is equal to $-\omega^2$, and the second order spatial derivative on the left side of (\ref{eq_a5}) is equal to $-k^2$. Since momentum is conserved across the time interface, the spatial derivative on the left side of (\ref{eq_a5}) is equivalent in both states, before and after the time interface.

Using the initial value theorem $\lim_{t\to0^{+}} f(t) = \lim_{s\to\infty} sF(s)$ \cite{Nilsson_2012} and (\ref{eq_a8}) for $V(x,s)$ above, and simplifying with $s_{1}=j\omega_{1}$ and $s_{2}=j\omega_{2}$, the boundary condition across the time boundary is obtained from,


\begin{align}
\lim_{t\to0^{+}} v(x,t) &= v(x,0^{+})\\ \label{eq_a11}
\lim_{s\to\infty} sV(x,s) &= \frac{s^2C_{1}(1 + \nicefrac{s_{1}}{s})}{s^2C_{2}(1 - (\nicefrac{s_{2}}{s})^2)} v(x,0^{-})
\end{align}

\noindent which yields the first boundary condition:

\begin{equation}
C_{1}v(x,0^{-}) = C_{2}v(x,0^{+}) \label{eq_a13}
\end{equation}

\hfill

This boundary condition state that the total charge $Q$ remains constant across the time boundary. Taking the time derivative $\frac{\partial}{\partial t}$ of this first boundary condition yields:

\begin{equation}
C_{1}\frac{\partial v(x,0^{-})}{\partial t} = C_{2}\frac{\partial v(x,0^{+})}{\partial t} \label{eq_a14}
\end{equation}

\hfill

\noindent which states that the current across the time boundary also remains constant. Thus we get the second boundary condition across the time boundary as:

\begin{equation}
I_{1} = I_{2} \label{eq_a15}
\end{equation}

\hfill

Finally, the reflection and transmission coefficients can be derived from the scattering equations before and after the switching event using the boundary conditions derived above. Without losing generality, the switching event is placed at time $t=0$. At $t<0$, before the switching event, the incident voltage and current are,

\begin{equation}
V_{1}=V_{01}\mathrm{e}^{s_{1}t-jkx} \label{eq_a16}
\end{equation}

\begin{equation}
I_{1}=I_{01}\mathrm{e}^{s_{1}t-jkx} = \frac{V_{01}}{Z_{1}}\mathrm{e}^{s_{1}t-jkx} \label{eq_a17}
\end{equation}

\hfill

\noindent where the voltage and current are related by the impedance in the initial state:

\begin{equation}
Z_{1} = \frac{V_{01}}{I_{01}} = \sqrt{\frac{L}{C_{1}}} \label{eq_a18}
\end{equation}

\hfill

At $t>0$, after the switching event, the voltage and current can be represented as:

\begin{equation}
V_{2}=[T\mathrm{e}^{s_{2}t} + R\mathrm{e}^{-s_{2}t}]V_{01}\mathrm{e}^{-jkx}\label{eq_a19}
\end{equation}

\begin{equation}
I_{2}=\frac{1}{Z_{2}}[T\mathrm{e}^{s_{2}t} - R\mathrm{e}^{-s_{2}t}]V_{01}\mathrm{e}^{-jkx}\label{eq_a20}
\end{equation}

\hfill

\noindent where $R$ is the reflection coefficient, and $T$ the transmission coefficient, and where the voltage and current are again related by the impedance, now in the second state:

\begin{equation}
Z_{2} = \frac{V_{02}}{I_{02}} = \sqrt{\frac{L}{C_{2}}} \label{eq_a21}
\end{equation}

\hfill

It is also noted that the transmitted signal and reflected signal have similar angular frequency magnitudes $\vert\omega_{2}\vert$, but are 180 degrees out of phase in order to have distinct fields upon the time discontinuity \cite{Mendonca_2002}. This is important when considering the phase relationship of the input, time reflected, and time refracting signals, and more specifically, the time reversal of the time-reflected wave.

Plugging the voltages and currents before and after the switching event into the boundary conditions (\ref{eq_a13}) and (\ref{eq_a15}) at $t=0$ and solving yields the following equations:

\begin{equation}
\frac{C_{1}}{C_{2}} = T + R \label{eq_a22}
\end{equation}

\begin{equation}
\frac{Z_{2}}{Z_{1}} = T - R \label{eq_a23}
\end{equation}

\hfill

From these equations the relationship for the reflected and transmitted coefficients in terms of either the capacitance's of the two states $C_{1}$ and $C_{2}$, or the impedances of the two states $Z_{1}$ or $Z_{2}$, can be derived as the following \cite{Moussa_2023}:

\begin{equation}
R = \frac{1}{2}\left[\frac{C_{1}}{C_{2}} - \sqrt{\frac{C_{1}}{C_{2}}}\right] = \frac{Z_{2}(Z_{2} - Z_{1})}{2Z_{1}^2} \label{eq_a24}
\end{equation}

\begin{equation}
T = \frac{1}{2}\left[\frac{C_{1}}{C_{2}} + \sqrt{\frac{C_{1}}{C_{2}}}\right] = \frac{Z_{2}(Z_{2} + Z_{1})}{2Z_{1}^2} \label{eq_a25}
\end{equation}

\hfill \\




\noindent \textbf{S2. Extraction of Time-Boundary Reflection and Transmission Coefficients}

\indent Due to both spatial-boundaries and time-boundaries within the experiment, the time-boundary reflection and transmission coefficients must be properly extracted from the measured and simulated responses at $V_\mathrm{in}$ and $V_\mathrm{out}$. Here we show the calculations and procedures used in order to extract their magnitudes from the measured and simulated data. This analysis assumes no insertion losses outside the microstrip line and power splitter, i.e., the losses associated with the coaxial cables, which for the frequency range of this work were deemed negligible.

First, the spatial-boundary reflection and transmission coefficients are extracted from the measured or simulated responses in State~1 (photodiodes in the OFF-state) and State~2 (photodiodes in the ON-state), both in steady-state without a time-boundary. From the response at $V_\mathrm{in}$, the peak voltage of the pulse at SB1, $V_\mathrm{SB1}$, as shown in Fig.~\ref{fig:no_kick_state_1}~(a) for State~1 and Fig.~\ref{fig:no_kick_state_2}~(a) for State~2, along with the input voltage $V_\mathrm{input}$, can be used to calculate the first spatial reflection $\Gamma_\mathrm{SB1}$,

\begin{equation}
V_\mathrm{SB1} = V_\mathrm{input}\,\Gamma_\mathrm{SB1}\,PS.
\end{equation}

\hfill

\noindent where $PS$ is the magnitude of the power splitter ratio including insertion loss. Assuming conservation of energy at a spatial boundary \cite{Pozar_2012}, the transmission coefficient can then be calculated from $T_\mathrm{SB1}=1+\Gamma_\mathrm{SB1}$.

Next, from the output voltage response $V_\mathrm{out}$ shown in Fig.~\ref{fig:no_kick_state_1}~(b) for State~1, with the peak voltage of the pulse at SB2, $V_\mathrm{SB2}$, the transmission coefficient at the second spatial-boundary $T_\mathrm{SB2}$ is extracted using the following equation:

\begin{equation}
V_\mathrm{SB2} = V_\mathrm{input}\,T_\mathrm{SB1}\,\mathrm{Loss}\,T_\mathrm{SB2}.
\end{equation}

\hfill

\noindent where $\mathrm{Loss}$ is the voltage ratio of the transmission power loss according to $\mathrm{Loss}=1- \lvert S_{11} \rvert ^2- \lvert S_{21} \rvert ^2$ \cite{Pozar_2012}, taken from the frequency response in Fig.~\ref{fig:freq_resp}. Similar to before, the reflection coefficient $\Gamma_\mathrm{SB2}$ is calculated using the conservation of energy as $\Gamma_\mathrm{SB2}=T_\mathrm{SB2}-1$ \cite{Pozar_2012}. For a Gaussian shaped pulse, a weighted average of the pulse in the frequency domain is used to calculate the transmission loss across the FWHM bandwidth of the pulse. 

For State~2, due to increased frequency dependent loss and dispersion visible in Fig.~\ref{fig:freq_resp}, the transmission loss is approximated to be $V_\mathrm{SB2}/V_\mathrm{input}$, yielding $T_\mathrm{SB2} \approx T_\mathrm{SB1}^{-1}$. Furthermore, for both states, the reflection and transmission coefficients at SB3 are assumed to be equivalent to those at SB2 due to symmetry.

\noindent With the spatial reflection and transmission coefficients known, the characteristic impedance of the microstrip line in both State~1 and State~2 is calculated using the well known equation $\Gamma=(Z_1 - Z_0)/(Z_1 + Z_0)$ \cite{Pozar_2012}.

As detailed in the main text, when the characteristic impedance of the microstrip line is changed from State~1 to State~2, a time-boundary is formed. If a signal is propagating within the microstrip line during this time-boundary, a time-scattering event occurs, generating a time-reflected and time-transmitted wave. Fig.~\ref{fig:signal_flow_with_kick} demonstrates the signal flow with a time-boundary generated by changing the impedance of the microstrip line from State~1 to State~2. From this, along with the simulated and measured responses at $V_\mathrm{in}$ and $V_\mathrm{out}$ and the previously derived spatial-boundary coefficients, we now outline the extraction process for the time-boundary reflection and transmission coefficients.

Beginning with the input response $V_\mathrm{in}$ for the time-boundary experiment shown in Fig.~\ref{fig:with_kick_single_pulse}~(a), the peak voltage magnitude $V_\mathrm{SB1}$ of the pulse at SB1 is used to extract the time reflection coefficient $\Gamma_\mathrm{TB1}$,

\begin{equation}
V_\mathrm{TB1} = V_\mathrm{input}\,T_\mathrm{SB1}^{(1)} \, \mathrm{Loss}^{(1)}_\text{1⁄2} \, \Gamma_\mathrm{TB1} \, \mathrm{Loss}^{(2)}_\text{1/2} \, T_\mathrm{SB3}^{(2)} \, PS
\end{equation}

\hfill

\noindent The superscript indicates which state the spatial-reflection coefficients and the transmission loss are in; before the time-boundary, the microstrip is in State~1, while after the boundary the microstrip is in State~2. The bandwidth of the pulse also changes between State~1 and State~2 due to frequency translation. Furthermore, we assume the signal has transmitted half the length of the microstrip line before the time event at $t = t_\mathrm{kick}$, thus only half the power loss is used for each state, as indicated by the $\mathrm{Loss}_{1/2}$ term. This is a reasonable assumption considering the microstrip line for this experiment was designed to be approximately the width of the pulse in the time-domain considering the phase velocity of State~1. The timing of the kick was carefully synchronized to occur right when the pulse has fully entered the line, and just before it exits, according to the discussion in Section~\ref{sec4}. The actual length of travel of the pulse can be calculated if more accuracy is required \cite{Moussa_2023}.

Finally, using the output response $V_\mathrm{out}$ for the time-boundary experiment shown in Fig.~\ref{fig:with_kick_single_pulse}~(b), the peak voltage magnitude $V_\mathrm{SB2}$ of the pulse at SB2 is used to extract the time transmission coefficient $T_\mathrm{TB1}$ using the following equation:

\begin{equation}
V_\mathrm{SB2} = V_\mathrm{input} \, T_\mathrm{SB1}^{(1)} \, \mathrm{Loss}^{(1)}_{1/2} \, T_\mathrm{TB1} \, \mathrm{Loss}^{(2)}_{1/2} \, T_\mathrm{SB2}^{(2)}
\end{equation}

\hfill

\noindent \textbf{S3. Chirped Pulse Parameters for Time Boundary Experiment with Phase Variation}

A chirped pulse with a linear instantaneous frequency $f(t)=ct+f_{0}$ has a quadratic instantaneous phase:

\begin{equation}
\phi(t) = 2\pi(\frac{c}{2}t^2 + f_0t) + \phi_{0} \label{secB1}
\end{equation}

\hfill

\noindent where $c$ is the chirpiness, defined as:

\begin{equation}
c=\frac{f_1 - f_0}{T} \label{secB2}
\end{equation}

\hfill

\noindent with $f_0$ the initial frequency, and $f_1$ the end frequency, of the chirp, and $T$ the transition time between $f_0$ to $f_1$. The initial phase at $t=0$ is represented by $\phi_{0}$. Considering a travelling wave with spatial representation in a non-dispersive medium, with $t \rightarrow t-x/\nu_\mathrm{p}$, the voltage of the chirped signal can then be represented as:

\begin{equation}
v(x,t) = cos(\pi c (t - \frac{x}{\nu_\mathrm{p}})^2 + \omega t - kx) \label{secB3}
\end{equation}

\hfill

\noindent where $\nu_\mathrm{p}$ is the phase velocity of the transmission line.







\end{document}